\documentclass[times,twocolumn]{aastex7}

\shorttitle{A systematic search for AGN flares}
\shortauthors{L. He et al.}

\usepackage{amsmath}
\usepackage{dashrule}
\usepackage{hyperref}
\usepackage{times}
\begin{document}
\title{A Systematic Search for Active Galactic Nucleus Flares in ZTF Data Release 23 }

\correspondingauthor{Lei He, Zheng-Yan Liu, Wen Zhao}

\author[0000-0001-7613-5815]{Lei He}
\affiliation{Department of Astronomy, University of Science and Technology of China, Hefei, Anhui 230026, China}
\affiliation{School of Astronomy and Space Sciences, University of Science and Technology of China, Hefei, Anhui 230026, China}
\email[show]{helei0831@mail.ustc.edu.cn}

\author[0000-0002-2242-1514]{Zheng-Yan Liu}
\affiliation{Department of Astronomy, University of Science and Technology of China, Hefei, Anhui 230026, China}
\affiliation{School of Astronomy and Space Sciences, University of Science and Technology of China, Hefei, Anhui 230026, China}
\email[show]{ustclzy@mail.ustc.edu.cn}

\author[0000-0001-9098-6800]{Rui Niu}
\affiliation{Department of Astronomy, University of Science and Technology of China, Hefei, Anhui 230026, China}
\affiliation{School of Astronomy and Space Sciences, University of Science and Technology of China, Hefei, Anhui 230026, China}
\email{nrui@ustc.edu.cn}

\author[0009-0002-5634-8842]{Ming-Shen Zhou}
\affiliation{Department of Astronomy, University of Science and Technology of China, Hefei, Anhui 230026, China}
\affiliation{School of Astronomy and Space Sciences, University of Science and Technology of China, Hefei, Anhui 230026, China}
\email{zmetassin@mail.ustc.edu.cn}

\author[0009-0000-5321-5775]{Pu-Run Zou}
\affiliation{Department of Astronomy, University of Science and Technology of China, Hefei, Anhui 230026, China}
\affiliation{School of Astronomy and Space Sciences, University of Science and Technology of China, Hefei, Anhui 230026, China}
\email{zpr2004@mail.ustc.edu.cn}

\author[0009-0002-0900-9901]{Bing-Zhou Gao}
\affiliation{Department of Astronomy, University of Science and Technology of China, Hefei, Anhui 230026, China}
\affiliation{School of Astronomy and Space Sciences, University of Science and Technology of China, Hefei, Anhui 230026, China}
\email{supernova_gbz@mail.ustc.edu.cn}

\author[0000-0001-6223-840X]{Run-Duo Liang}
\affiliation{National Astronomical Observatories, Chinese Academy of Sciences, 20A Datun Road, Beijing 100101, China}
\email{liangrd@bao.ac.cn}

\author[0000-0001-7688-6504]{Liang-Gui Zhu}
\affiliation{Department of Astronomy, University of Science and Technology of China, Hefei, Anhui 230026, China}
\affiliation{School of Astronomy and Space Sciences, University of Science and Technology of China, Hefei, Anhui 230026, China}
\email{lianggui.zhu@pku.edu.cn}

\author[0000-0001-9449-9268]{Jian-Min Wang}
\affiliation{Key Laboratory for Particle Astrophysics, Institute of High Energy Physics, Chinese Academy of Sciences, 19B Yuquan Road, Beijing 100049, China}
\email{wangjm@ihep.ac.cn}

\author[0000-0002-7152-3621]{Ning Jiang}
\affiliation{Department of Astronomy, University of Science and Technology of China, Hefei, Anhui 230026, China}
\affiliation{School of Astronomy and Space Sciences, University of Science and Technology of China, Hefei, Anhui 230026, China}
\email{jnac@ustc.edu.cn}

\author[0000-0002-4223-2198]{Zhen-Yi Cai}
\affiliation{Department of Astronomy, University of Science and Technology of China, Hefei, Anhui 230026, China}
\affiliation{School of Astronomy and Space Sciences, University of Science and Technology of China, Hefei, Anhui 230026, China}
\email{zcai@ustc.edu.cn}

\author[0000-0002-9092-0593]{Ji-an Jiang}
\affiliation{Department of Astronomy, University of Science and Technology of China, Hefei, Anhui 230026, China}
\affiliation{School of Astronomy and Space Sciences, University of Science and Technology of China, Hefei, Anhui 230026, China}
\email{jian.jiang@ustc.edu.cn}

\author[0000-0002-7835-8585]{Zi-Gao Dai}
\affiliation{Department of Astronomy, University of Science and Technology of China, Hefei, Anhui 230026, China}
\affiliation{School of Astronomy and Space Sciences, University of Science and Technology of China, Hefei, Anhui 230026, China}
\email{daizg@ustc.edu.cn}

\author[0000-0002-7330-4756]{Ye-Fei Yuan}
\affiliation{Department of Astronomy, University of Science and Technology of China, Hefei, Anhui 230026, China}
\affiliation{School of Astronomy and Space Sciences, University of Science and Technology of China, Hefei, Anhui 230026, China}
\email{yfyuan@ustc.edu.cn}

\author[0000-0003-4280-7673]{Yong-Jie Chen}
\affiliation{Key Laboratory for Particle Astrophysics, Institute of High Energy Physics, Chinese Academy of Sciences, 19B Yuquan Road, Beijing 100049, China}
\email{chenyj20@outlook.com}

\author[0000-0002-1330-2329]{Wen Zhao}
\affiliation{Department of Astronomy, University of Science and Technology of China, Hefei, Anhui 230026, China}
\affiliation{School of Astronomy and Space Sciences, University of Science and Technology of China, Hefei, Anhui 230026, China}
\affiliation{College of Physics, Guizhou University, Guiyang, 550025, China}
\email[show]{\\ wzhao7@ustc.edu.cn}

\begin{abstract}
  Active galactic nuclei (AGNs) are known to exhibit stochastic variability across a wide range of timescales and wavelengths. AGN flares are extreme outbursts that deviate from this typical behavior and may trace a range of energetic physical processes. Using six years of data from Zwicky Transient Facility (ZTF) Data Release 23, we conduct a systematic search for AGN flares among a sample of well-sampled AGNs and AGN candidates. We construct two catalogs: the AGN Flare Coarse Catalog (AGNFCC), containing 28,504 flares identified via Bayesian blocks and Gaussian Processes, and the AGN Flare Refined Catalog (AGNFRC), comprising 1,984 high-confidence flares selected using stricter criteria. We analyze their spatial distribution, temporal characteristics, host AGN type and potential origins. Some flares can be associated with known supernovae, tidal disruption events, or blazars, and a few may be linked to binary black hole mergers or microlensing events. These catalogs provide a valuable resource for studying transient phenomena in AGNs and are publicly available at \url{https://github.com/Lyle0831/AGN-Flares}.

\end{abstract}


\section{Introduction}
Active galactic nuclei (AGNs) are extremely luminous sources powered by accretion onto supermassive black holes (SMBHs), and they exhibit significant variability across different wavelengths \citep{ulrichVARIABILITYACTIVEGALACTIC1997, kellyAREVARIATIONSQUASAR2009}. The optical light curves of AGNs often show stochastic brightness fluctuations on timescales from days to years, commonly modeled by a one-dimensional damped random walk \citep{macleodMODELINGTIMEVARIABILITY2010,kozlowskiRevisitingStochasticVariability2016}. These fluctuations can be attributed to various processes occurring near the supermassive black hole, such as accretion disk instabilities \citep{treveseQuasarSpectralSlope2002}, disk inhomogeneities \citep{dexterQUASARACCRETIONDISKS2010,caiSimulatingTimescaleDependentColor2016,caiEUCLIAExploringUV2018}, or X-ray reprocessing  \citep{kubotaPhysicalModelBroadband2018}, but the primary origin is still under debate. In addition to the general optical variability of AGN over different timescales, a subset of these systems exhibit transient and extreme events, known as AGN flares \citep{grahamUnderstandingExtremeQuasar2017,lawrenceSlowblueNuclearHypervariables2016}. These events represent a significant departure from the typical stochastic variability of AGN, yet their origins remain largely uncertain.

A growing body of theoretical work suggests that many of these extreme flaring events may originate from stellar and compact objects embedded in AGN accretion disks, broadly referred to as accretion-modified stars \citep[AMSs,][]{wangAccretionmodifiedStarsAccretion2021a, wangAccretionmodifiedStarsAccretion2021, liuAccretionmodifiedStarsAccretion2024}. Within such dense and gas-rich environments, stars may evolve rapidly and end their lives as supernovae \citep{liCorecollapseSupernovaExplosions2023}, or be tidally disrupted by the central SMBH \citep{ryuTidalDisruptionsMainsequence2020} or stellar-mass black holes \citep{yangTidalDisruptionStellarmass2022}. These processes contribute to the metal enrichment of the AGN disk \citep{wangStarFormationSelfgravitating2023} and the formation of compact objects. A large population of compact objects can form binaries \citep{rowanBlackHoleBinary2023,wangStellarBHPopulation2024} and undergo mergers \citep{mckernanBlackHoleNeutron2020}. The resulting coalescences can produce gravitational waves and potentially be accompanied by observable electromagnetic signals, including transients from white dwarf collisions \citep{zhangElectromagneticSignaturesWhite2023}, kilonova emission from neutron star-neutron star or neutron star-black hole mergers \citep{renInteractingKilonovaeLonglasting2022}, and flares from binary black hole mergers \citep{grahamCandidateElectromagneticCounterpart2020,grahamLightDarkSearching2023}

These scenarios offer a unified framework for interpreting diverse AGN flaring phenomena and highlight the importance of searching for AGN flares as a means to probe the AMS populations. In addition, AGN flares serve as valuable diagnostics of accretion physics. Their timescales and amplitudes may constrain the properties of AGN disks and act as important probes for accretion rates \citep{ruanUNDERSTANDINGCHANGINGLOOKQUASARS2016, grahamUnderstandingExtremeQuasar2017}. Multi-epoch spectroscopy during the period of the flares further enables reverberation mapping to get the location of the emission region, helping constrain the structure of the AGN disk \citep{payneChandraHSTSTIS2023,cabreraSearchingElectromagneticEmission2024a}.

Recognizing their scientific potential, several systematic searches have been conducted to identify AGN flares \citep{lawrenceSlowblueNuclearHypervariables2016,grahamUnderstandingExtremeQuasar2017}, utilizing data from large surveys like Panoramic Survey Telescope and Rapid Response System \citep[Pan-STARRS,][]{kaiserPanSTARRSWidefieldOptical2010}, Sloan Digital Sky Survey \citep[SDSS,][]{sesarExploringVariableSky2007}, and Catalina Real-time Transient Survey \citep[CRTS,][]{drakeFirstResultsCatalina2009}. These studies have primarily focused on detecting long-duration, high-amplitude variability events, primarily due to the limited observational cadence of these surveys.

The emergence of high-cadence time-domain facilities, such as the Zwicky Transient Facility \citep[ZTF,][]{bellmZwickyTransientFacility2019,grahamZwickyTransientFacility2019}, the Wide Field Survey Telescope \citep[WFST,][]{wangScience25meterWide2023}, and the Vera C. Rubin observatory \citep{collaborationScienceDrivenOptimizationLSST2017,ivezicLSSTScienceDrivers2019}, would be able to capture much more AGN flares across a wide range of timescales. These advances would facilitate the discovery and characterization of AGN flares, offering deeper insights into the physical mechanisms driving such outbursts.

To fully utilize the data from modern high-cadence surveys, recent studies have proposed the use of Gaussian Processes (GPs) for the detection of AGN flares \citep{mclaughlinUsingGaussianProcesses2024}. \citet{grahamLightDarkSearching2023} applied a GP-based technique to separate flaring events from typical AGN variability, while \citet{heTracingLightIdentification2025} further demonstrated that applying such methods to larger datasets significantly improves the reliability of flare identification. Motivated by these developments, we carry out a systematic search for AGN flares in ZTF Data Release 23 (DR23) using a GP-based approach, aiming to construct the most comprehensive catalog of such events to date.

The outline of this paper is as follows. In Section \ref{sec:data}, we describe the ZTF dataset and the AGN catalogs used in this work. In Section \ref{sec:method}, we introduce the flare detection method, including Gaussian Process modeling and the selection criteria used to construct two catalogs: the AGN Flare Coarse Catalog (AGNFCC) and the AGN Flare Refined Catalog (AGNFRC). In Section \ref{sec:catalog}, we present the statistical properties of the detected flares. In Section \ref{sec:discussion}, we discuss their potential origins. Finally, we summarize our main findings in Section \ref{sec:summary}.

\section{data \label{sec:data}}
\subsection{ZTF DR23}
The Zwicky Transient Facility (ZTF) is a wide-field time-domain survey utilizing a 47 deg$^{2}$ field-of-view camera mounted on the Palomar 48-inch Samuel Oschin Schmidt telescope. Designed to systematically explore transient and variable phenomena with high cadence and sensitivity \citep{bellmZwickyTransientFacility2018,grahamZwickyTransientFacility2019}, ZTF has been operating since March 2018, covering approximately $3\pi$ sr across the Northern sky \citep{bellmZwickyTransientFacility2019}. For the Northern sky of declination $>-31^{\circ}$, ZTF operates two primary observing programs: the \added{Galactic Plane} Survey, which targets fields with $|b|<7^{\circ}$ at a 1-day cadence, and the Northern Sky Survey, which observes fields with  $|b|>7^{\circ}$ at a 3-day cadence. Both surveys acquire data in the g and r bands, providing dual-band light curves for each source.

Our analysis is based on ZTF Data Release 23 (DR23), publicly released on January 21, 2025. This release consolidates all observations from public surveys up to 2024 October 31 and private surveys prior to 2023 June 30, providing one of the most comprehensive time-domain datasets to date for the northern sky.

\subsection{AGN Catalogs}

To extract AGN light curves from ZTF DR23, we need to positionally cross-match ZTF sources with pre-existing AGN catalogs that provide reliable classifications and positions. All the catalogs we used in this work are shown as follows.

\textit{Million Quasars Catalog} \citep[Milliquas, v8,][]{fleschMillionQuasarsMilliquas2023}: A comprehensive catalog of spectroscopically confirmed AGNs and high-confidence AGN candidates ($p_{\mathrm{QSO}} \ge 99\%$), compiled from all published papers up to 2023 June 30. It contains 1,021,800 sources, with $> 90\%$ having spectroscopic redshifts and the remainder utilizing photometric redshifts. This catalog incorporates astrometric corrections to historical positional errors and integrates high-precision multi-survey data, delivering the most precise locations.

\textit{DESI AGN Catalog} \citep{collaborationDataRelease12025}: The Dark Energy Spectroscopic Instrument (DESI) has conducted the most extensive extragalactic spectroscopic redshift survey to date. Its first Data Release (DR1) provides high-confidence redshifts for approximately 18 million objects across over 9,000 deg$^{2}$. In this work, we use the subset of objects labeled with \texttt{SPECTYPE = `QSO'} in the DESI value-added catalog\footnote{\url{https://data.desi.lbl.gov/public/dr1/vac/dr1/agnqso}} (Juneau et al 2025, in preparation), which includes a total of 1,772,694 AGNs. This represents the largest spectroscopic AGN sample available to date, making it a valuable source for statistical studies of AGN.

\textit{LAMOST AGN Catalog} \citep{aiLargeSkyArea2016, dongLargeSkyArea2018, yaoLargeSkyArea2019, jinLargeSkyArea2023}: A set of spectroscopically confirmed AGNs from the Large Sky Area Multi-Object Fiber Spectroscopic Telescope (LAMOST) quasar survey conducted between 2011 and 2021. The full sample contains 55,636 quasars, of which over 30,000 are independently discovered. All spectra are recalibrated using SDSS or Pan-STARRS photometry, and the key spectral properties, including emission line widths, continuum luminosities, and black hole masses, are homogeneously measured.

\textit{Quaia AGN Catalog} \citep{storey-fisherQuaiaGaiaunWISEQuasar2024}: An all-sky catalog constructed from sources identified as quasar candidates in Gaia Data Release 3 \citep{bailer-jonesGaiaDataRelease2023}, initially comprising 6,649,162 candidates with redshift estimates derived from the low-resolution blue photometer/red photometer spectra. To improve the purity of the sample, these candidates are cross-matched with the unWISE infrared catalog \citep{schlaflyUnWISECatalogTwo2019} and refined by applying cuts based on color and proper motion to reduce contamination. Redshift estimates are subsequently improved using a k-Nearest Neighbors model trained on spectroscopically confirmed quasars. The final catalog contains 1,295,502 quasar candidates with Gaia G band magnitudes brighter than 20.5.

\textit{WISE AGN R90 Catalog} \citep{assefWISEAGNCatalog2018}: A mid-infrared selected catalog of AGNs identified from the Wide-field Infrared Survey Explorer AllWISE Data Release\footnote{\url{http://wise2.ipac.caltech.edu/docs/release/allwise/}}. AGN candidates are selected using stringent color cuts in the W1 ($3.4\mu$m) and W2 ($4.6\mu$m) bands, which are optimized for 90\% reliability. Contaminants are mitigated through spatial filtering (e.g., Galactic Plane exclusion) and artifact rejection, yielding 4,543,530 AGNs across 30,093 deg$^{2}$ of extragalactic sky. This represents one of the largest AGN samples available and provides a foundational resource for systematic studies of AGN.

\begin{figure}[htb]
  \includegraphics[width=0.45\textwidth]{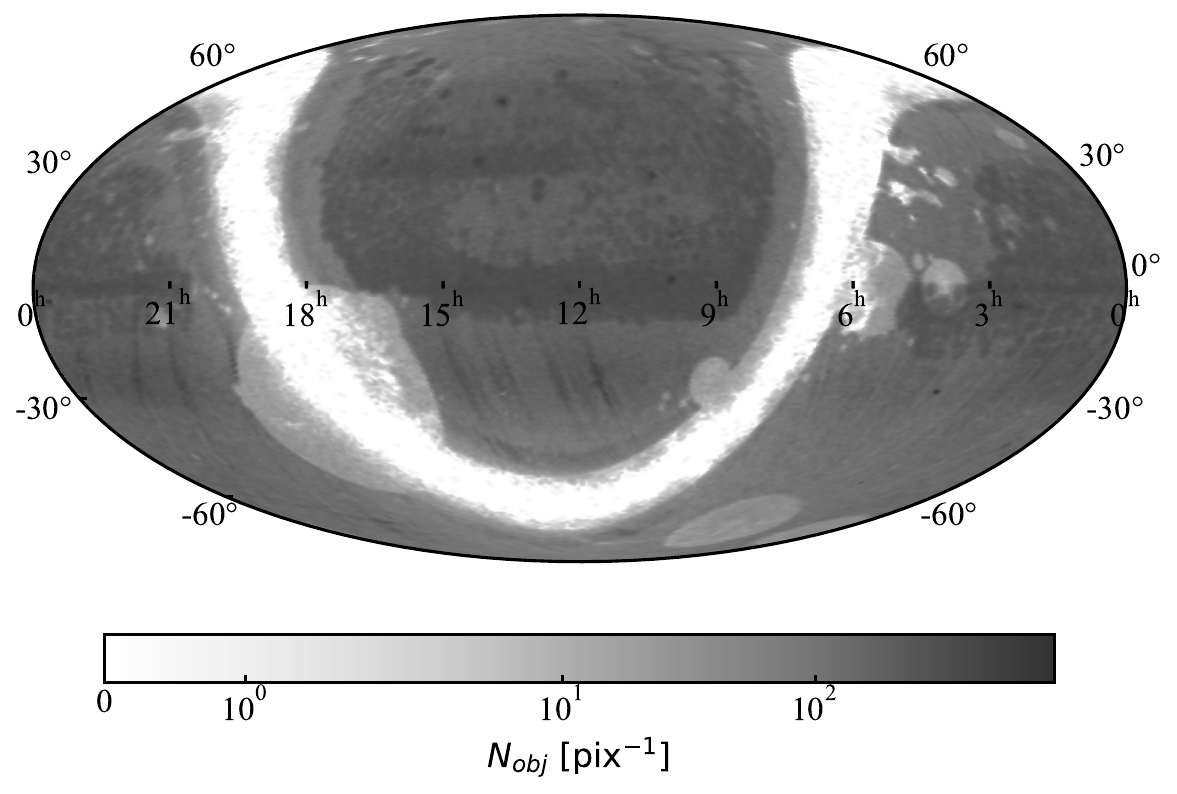}
  \caption{Mollweide projection of the sky positions in equatorial coordinates of the objects used in this work. The resolution is that of a HealPix map with NSIDE=64.}
  \label{fig:AGNcatalog}
\end{figure}

\begin{figure}[htb]
  \includegraphics[width=0.45\textwidth]{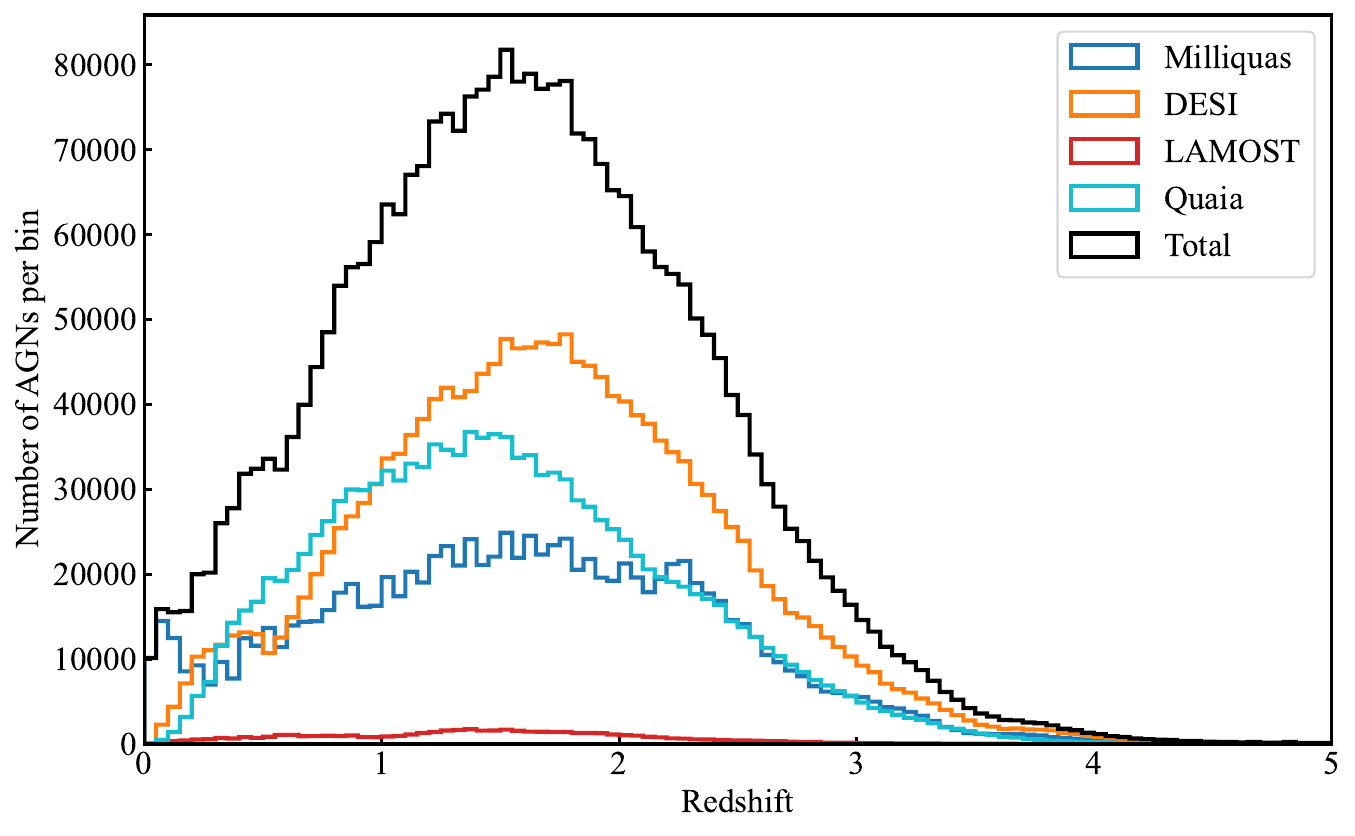}
  \caption{Redshift distributions of AGNs from different catalogs, binned with a bin size of $\Delta z=0.05$.}
  \label{fig:AGNRedshift}
\end{figure}

To remove duplicate entries across the various AGN catalogs, we perform a cross-match of all sources using a $1^{\prime\prime}$ radius. After eliminating overlaps, the final combined AGN sample contains 6,602,925 unique sources. Figure \ref{fig:AGNcatalog} shows the spatial distribution of these AGNs, where the color of each pixel reflects the number of objects in that region of the sky, and each pixel covers an area of approximately 0.84 square degrees. Figure \ref{fig:AGNRedshift} presents the redshift distributions of AGNs from different catalogs.

\subsection{Pre-processing}

We retrieve all light curves within the ZTF DR23 dataset using the bulk download option described in Section 12.c of the ZTF DR23 document\footnote{\url{https://irsa.ipac.caltech.edu/data/ZTF/docs/releases/dr23/ztf\_release\_notes\_dr23.pdf}}. \added{These light curves are generated from positionally matched sources detected across epochs using the PSF-fit photometry catalogs, with the reference image catalog serving as the seed for position matching \citep{masciZwickyTransientFacility2018}.} Each light curve contains the Heliocentric-based Modified Julian Date (HMJD), the point spread function (PSF) magnitude, 1-$\sigma$ photometric uncertainty, and corresponding photometric/image quality flags (\texttt{catflags}) for every observation. ZTF assigns unique object identifiers to sources based on combinations of right ascension (RA), declination (DEC), filter, field, and charge-coupled device (CCD) quadrant. We focus on the g and r bands data for their regular cadence and sky coverage \citep{bellmZwickyTransientFacility2019}, excluding the i band due to its insufficient sampling and limited source coverage.

\begin{figure}[htb]
 \includegraphics[width=0.45\textwidth]{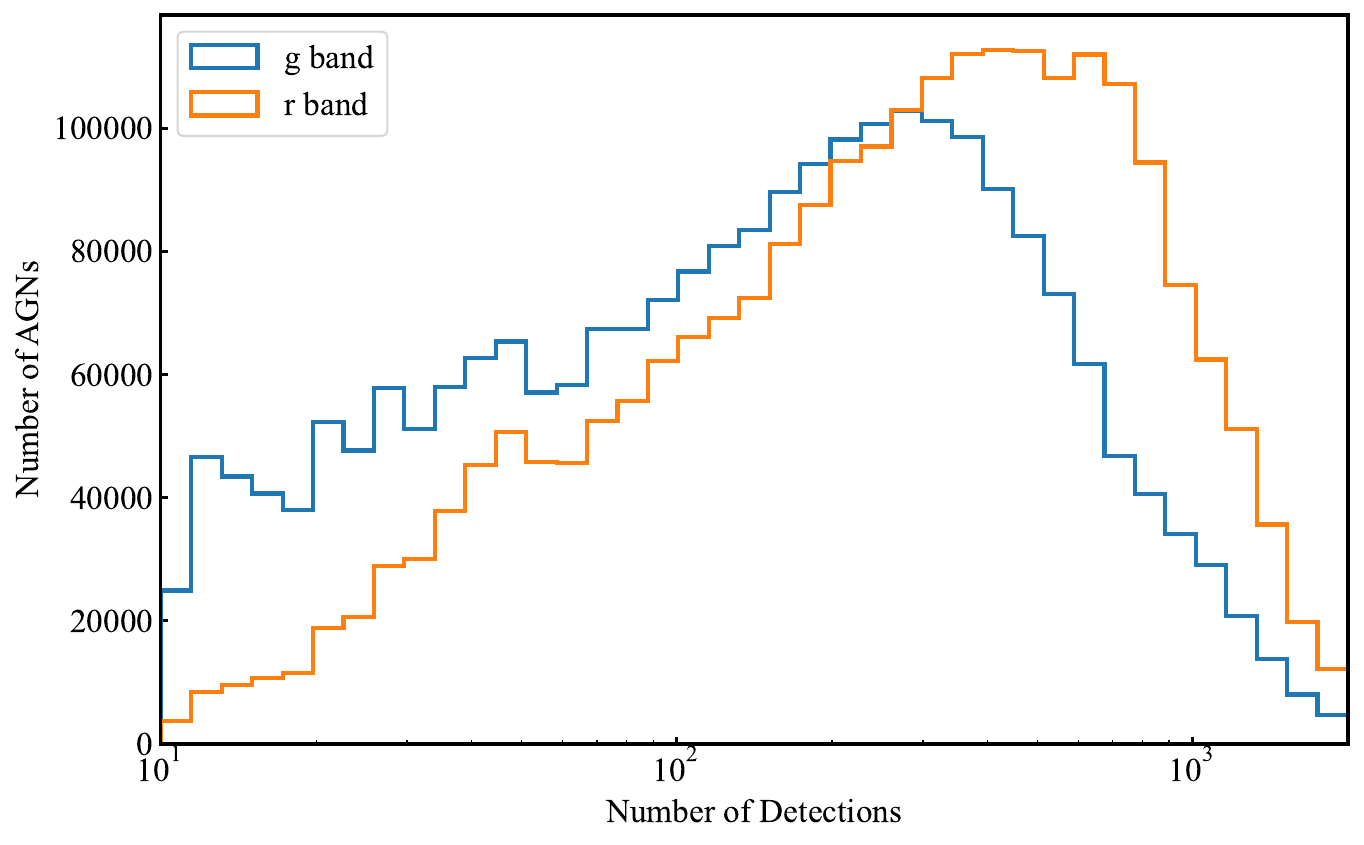}
  \caption{ZTF detection counts distribution for all AGNs. }
  \label{fig:AGNdetection}
\end{figure}

Similar to \citet{grahamLightDarkSearching2023}, we cross-match AGN positions from the combined catalog with ZTF DR23 sources using a $3^{\prime\prime}$ matching radius, resulting in approximately 3.6 million AGN light curves. To ensure sufficient sampling and data quality for flare detection, we require each light curve to contain more than 30 observations per band after removing epochs with \texttt{catflags} $>0$. This threshold is conservatively chosen given ZTF's six-year baseline and is intended to provide sufficient temporal coverage for distinguishing flares from the AGN intrinsic variability. Figure \ref{fig:AGNdetection} presents the distribution of valid ZTF detection counts per AGN in the g and r bands. Most AGNs have several hundred observations per band, ensuring robust coverage for variability analysis. The number of sources in each AGN catalog is shown in Table \ref{tab:SourceNumber}.

\begin{deluxetable*}{ccccccc}[htb]
  \tablecaption{The number of sources in each AGN catalog}
  \tablehead{
    \colhead{Vetting filters}& \colhead{Milliquas} & \colhead{DESI} & \colhead{LAMOST} & \colhead{Quaia} & \colhead{WISE R90} & \colhead{Total}
  }
  \label{tab:SourceNumber}
  \startdata
  initial AGNs & 1,021,800 & 1,772,694 & 55,636 & 1,295,502 & 4,543,530 & 6,602,925 \\
  ZTF cross-match & 950,298 & 1,437,122 & 53,143 & 937,883 & 1,969,553 &3,574,767 \\
  observation threshold & 721,553 & 778,663 & 49,484 & 850,747 & 1,061,365 & 1,966,641 \\
  \multicolumn{7}{c}{\hdashrule[0.5ex]{14cm}{1pt}{3pt}} \\
  potential flares & 224,247 & 177,766 & 30,133 & 351,476 & 319,689 & 510,511 \\
  AGNFCC & 11,779 & 8,237 & 991 & 14,999 & 16,729 & 28,504 \\
  AGNFRC & 944 & 513 & 64 & 1,128 & 1,128 & 1,984
  \enddata
\end{deluxetable*}

We employ a hybrid sigma-clipping algorithm to suppress noise while preserving astrophysical signals. Points deviating $> 3\sigma$ from the median flux are flagged as outliers, and single outliers are removed, whereas $> 2$ consecutive outliers or isolated outliers surrounded by points $>2.5\sigma$ are retained as potential flare signatures. Finally, cleaned light curves are binned into 3-day intervals to enhance the signal-to-noise ratio (SNR) for flare detection, using the inverse-variance weighting based on the photometric errors when computing the average flux in each bin.

\section{method\label{sec:method}}
\subsection{Gaussian Processes}
\added{We model the intrinsic AGN variability using Gaussian Process Regression (GPR), which is an interpolation method based on Gaussian Process (GP) governed by prior covariances \citep{chilesGeostatisticsModelingSpatial2012a}. Formally, a GP is a collection of random variables with a Gaussian joint distribution \citep{rasmussenGaussianProcessesMachine2006}.} GPs characterize the covariance structure of data through a mean function and a kernel, which encapsulates prior assumptions about the system's behavior, such as smoothness, periodicity, or multi-scale correlations.

In this work, we adopt the median of the light curve as the mean function, which is robust to outliers and typical stochastic AGN variability. As for the covariance function, we utilize the Matern-1/2 function:
\begin{equation}
  \label{eq:kernel}
k(t,t') = \rho^{2}\exp\left(-\frac{|t-t'|}{\tau}\right),
\end{equation}
where $\rho$ (variability amplitude) and $\tau$ (characteristic timescale) are hyperparameters. This kernel has proven to be effective in modeling AGN light curves \citep{griffithsModelingMultiwavelengthVariability2021} and it is computationally friendly.

Given $N$ observations $\boldsymbol{y} = \{y_{i}\}_{i=1,...,N}$ with uncertainties $\boldsymbol{\sigma} = \{\sigma_{i}\}$ at times $\boldsymbol{t}=\{t_{i}\}$, the covariance matrix becomes:
\begin{equation}
  K_{ij}=k(t_{i},t_{j})+\delta_{ij}\sigma_{i}^{2},
\end{equation}
where $\delta_{ij}$ is the discrete Kronecker delta function. The marginal log likelihood of observed data is:
\begin{equation}
  \log \mathcal{L} = -\frac12(\boldsymbol{y}-\boldsymbol{m})^{T}\boldsymbol{K}^{-1}(\boldsymbol{y}-\boldsymbol{m})-\frac12 \log |\boldsymbol{K}| - \frac N2\log 2\pi,
\end{equation}
where the vector $\boldsymbol{m}$ represents the mean function.

For a given AGN light curve, the hyperparameters $\{\rho,\tau\}$ in the GP model can be optimized via maximum likelihood estimation. In \citet{mclaughlinUsingGaussianProcesses2024}, GPs are employed to detect AGN flares by comparing the hyperparameters of individual AGNs to the overall hyperparameter distribution of the AGN population, with deviations potentially indicating the presence of a flare. However, since short-duration flares exert minimal influence on the parameter fits, this method is primarily sensitive to long-duration flares. \added{In contrast, \citet{grahamLightDarkSearching2023} proposed a change point detection method, in which a GP model is first fitted to the AGN light curve, and any variability within a specific time window that significantly deviates from this fitted model is identified as a potential flare.} In this work, we extend the approach from \citet{grahamLightDarkSearching2023} to detect flares of all durations across a large AGN sample.

If a flare occurs within a short segment of the light curve, then the hyperparameters $\rho_{0}$ and $\tau_{0}$ obtained from the whole light curve no longer adequately capture the variability during the flare. This inadequacy arises because the GP mode, trained on the baseline data, is not designed to account for the additional features introduced by the flare. To assess the deviation, a test statistic analogous to it is defined as follows:
\begin{equation}
  \label{eq:lambda}
  \lambda = (\boldsymbol{y}_{\mathrm{flare}}-\boldsymbol{m}_{0})^{T}\boldsymbol{K}_{0}^{-1}(\boldsymbol{y}_{\mathrm{flare}}-\boldsymbol{m}_{0}),
\end{equation}
where $\boldsymbol{y}_{\mathrm{flare}}$ represents the observations during the flare and $\boldsymbol{m}_{0}$ is determined as the median of the light curve.

\subsection{Initial Detection of AGN Flares \label{sec:flare_detection}}

For each AGN with available ZTF data, we first fit its light curve (in flux space) using a Bayesian block representation \citep{scargleSTUDIESASTRONOMICALTIME2013}. This technique optimally segments the data into a series of discontinuous, piecewise constant components, facilitating the detection of significant local variations. To identify potential flares, we apply the hill-climbing procedure proposed by \citet{meyerCharacterizingGammaRayVariability2019}. In this approach, peaks are identified as blocks that are higher than both the preceding and succeeding blocks, and these peaks are extended in both directions as long as the subsequent blocks remain lower. We define a flare interval in such a set of blocks as the segment where the flux exceeds 0.2 times the difference between the peak flux and the median flux of the entire light curve. If a flare interval consists of only one data point, we exclude it from further analysis.

Then we fit the entire light curve for each AGN using the kernel in Equation (\ref{eq:kernel}) to obtain the maximum likelihood hyperparameters \{$\rho_0, \tau_0$\} and compute the test statistic $\lambda$ for each flare interval following Equation (\ref{eq:lambda}). \added{To assess the significance of each potential flare, we generate a large number of simulated light curves using the derived \{$\rho_0,\tau_0$\} parameters, while preserving the exact cadence and uncertainties of the observed data.} For each simulations, we compute the corresponding $\lambda$ values for the flare intervals. Finally, by comparing the observed $\lambda$ value with the distribution of $\lambda$ values from the simulated light curves, we determine the significance of each potential flare. We define the significance as $p_{\mathrm{flare}}$, which represents the probability that the observed flare is not attributed to the intrinsic variability of the AGN.

\begin{figure}
  \centering
  \includegraphics[width=0.45\textwidth]{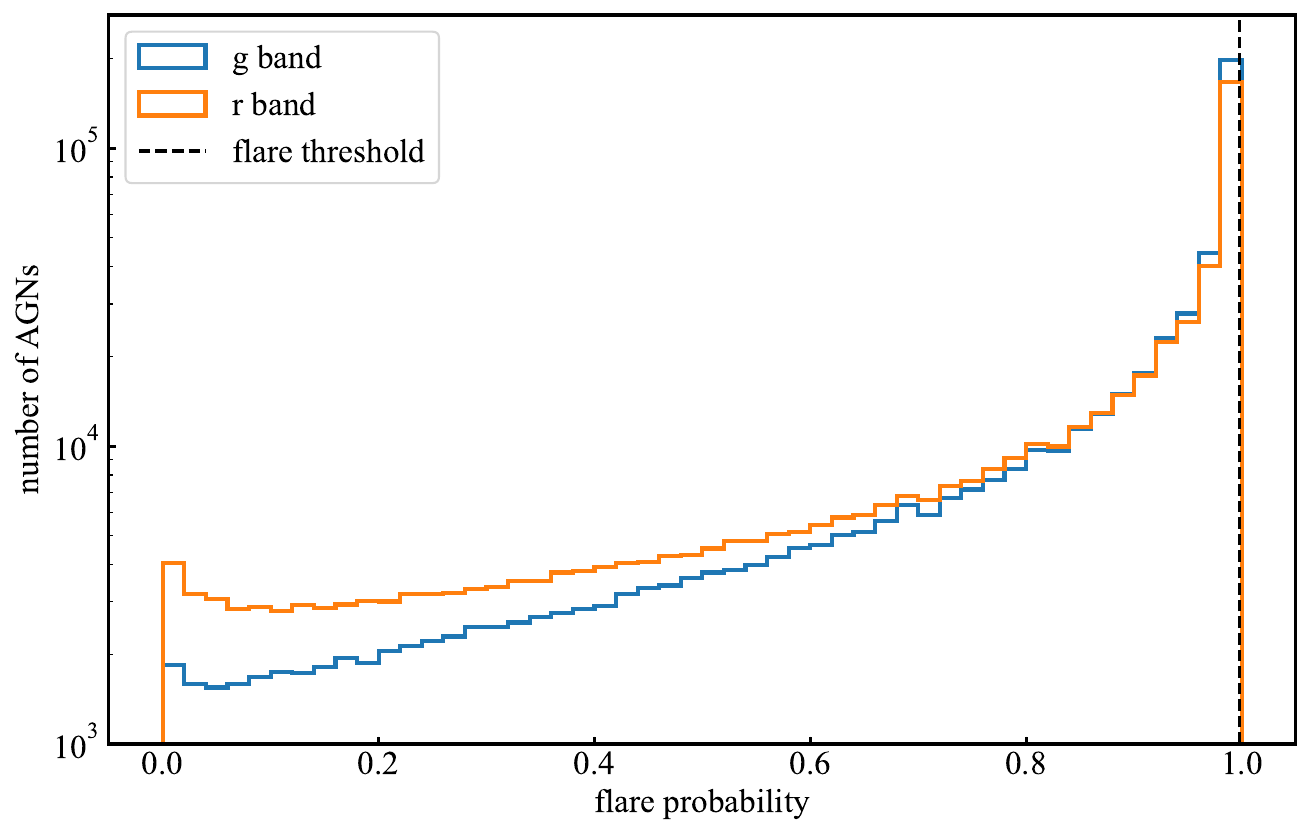}
  \caption{Distribution of the highest $p_{\mathrm{flare}}$ values for each AGN. The vertical dashed line represents the threshold of $p_{\mathrm{flare}}=0.998$.}
  \label{fig:FlareProb}
\end{figure}

Furthermore, a potential flare is considered genuine only if it is detected in both the g and r bands, and the overall $p_{\mathrm{flare}}$ for this potential flare is taken as the minimum of the values obtained from the two bands. Figure \ref{fig:FlareProb} shows the distribution of the highest $p_{\mathrm{flare}}$ value for each AGN. Given the limited statistical resolution inherent to our simulation-based approach, we adopt a threshold of $p_{\mathrm{flare}} > 0.998$ as a practical compromise between statistical reliability and computational efficiency. This threshold is intended to select only the most significant flare candidates, while allowing a small buffer to avoid missing genuine flare events.

\begin{figure*}[htb!]
  \centering
  \gridline{\fig{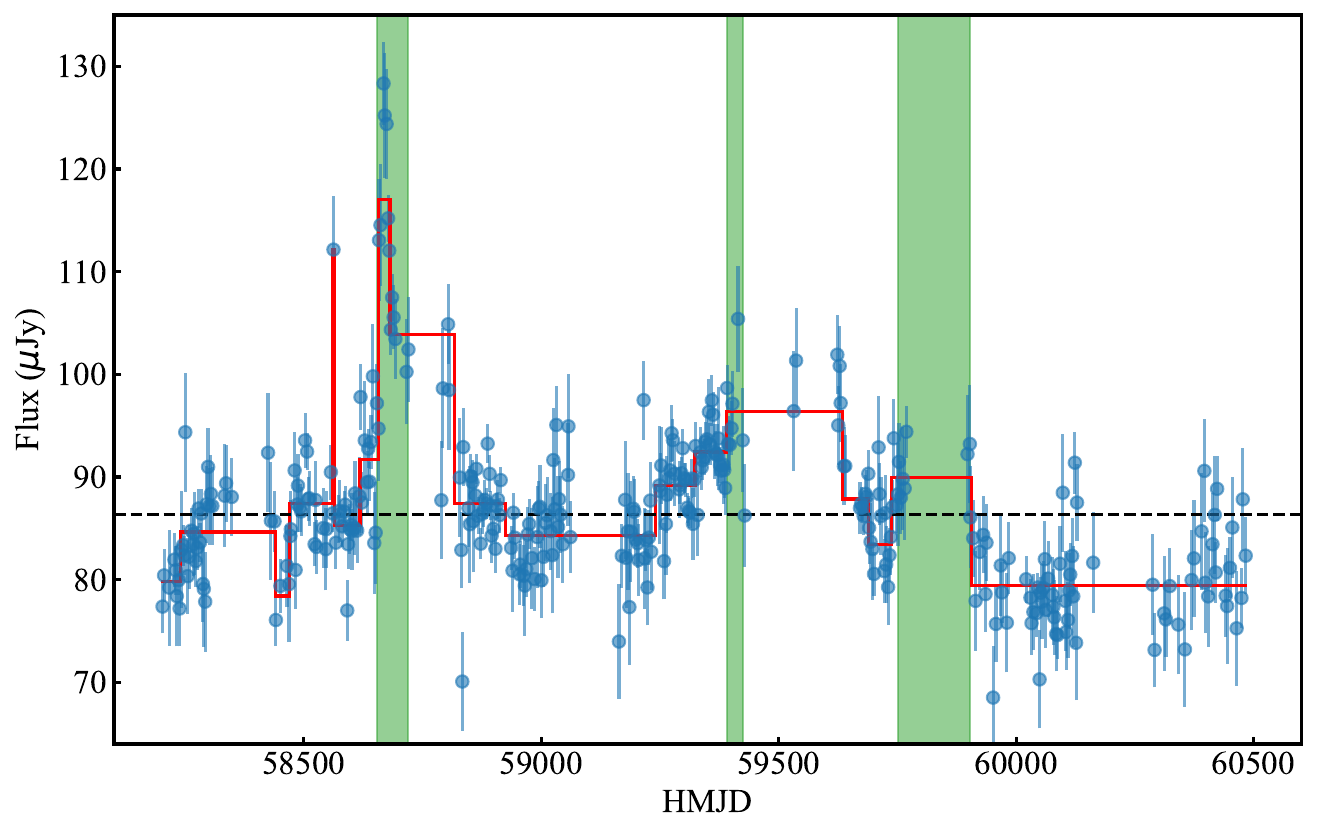}{0.5\textwidth}{(a)}
            \fig{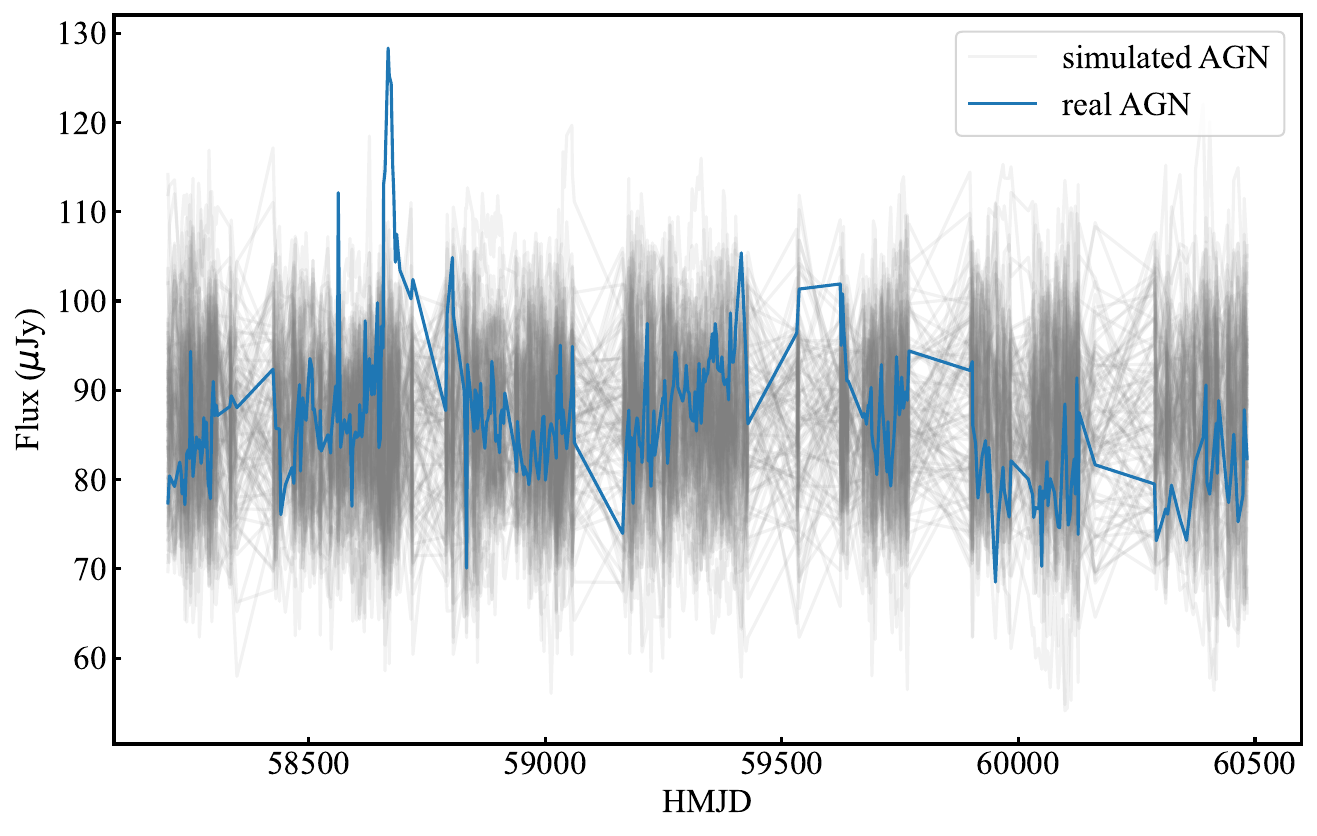}{0.5\textwidth}{(b)}
  }
  \gridline{\fig{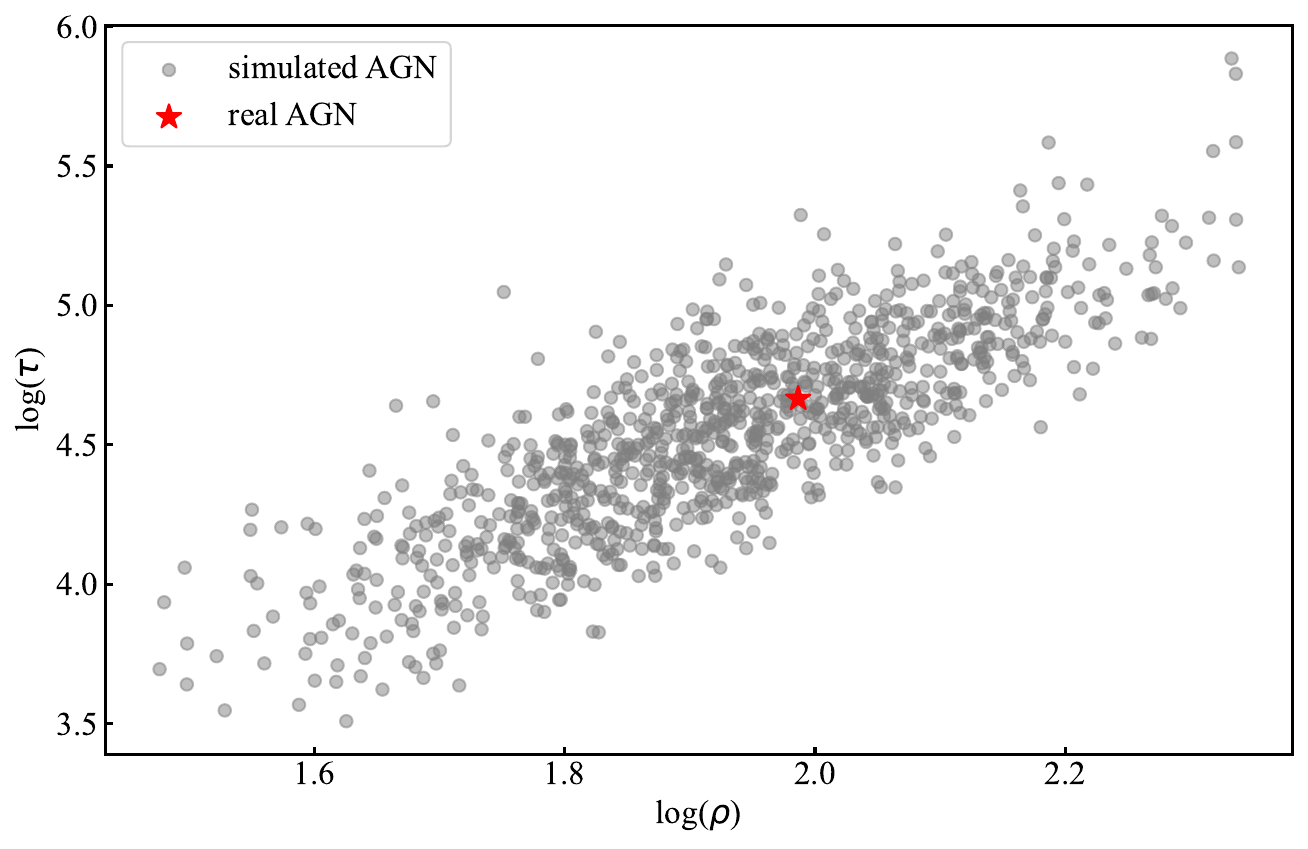}{0.5\textwidth}{(c)}
            \fig{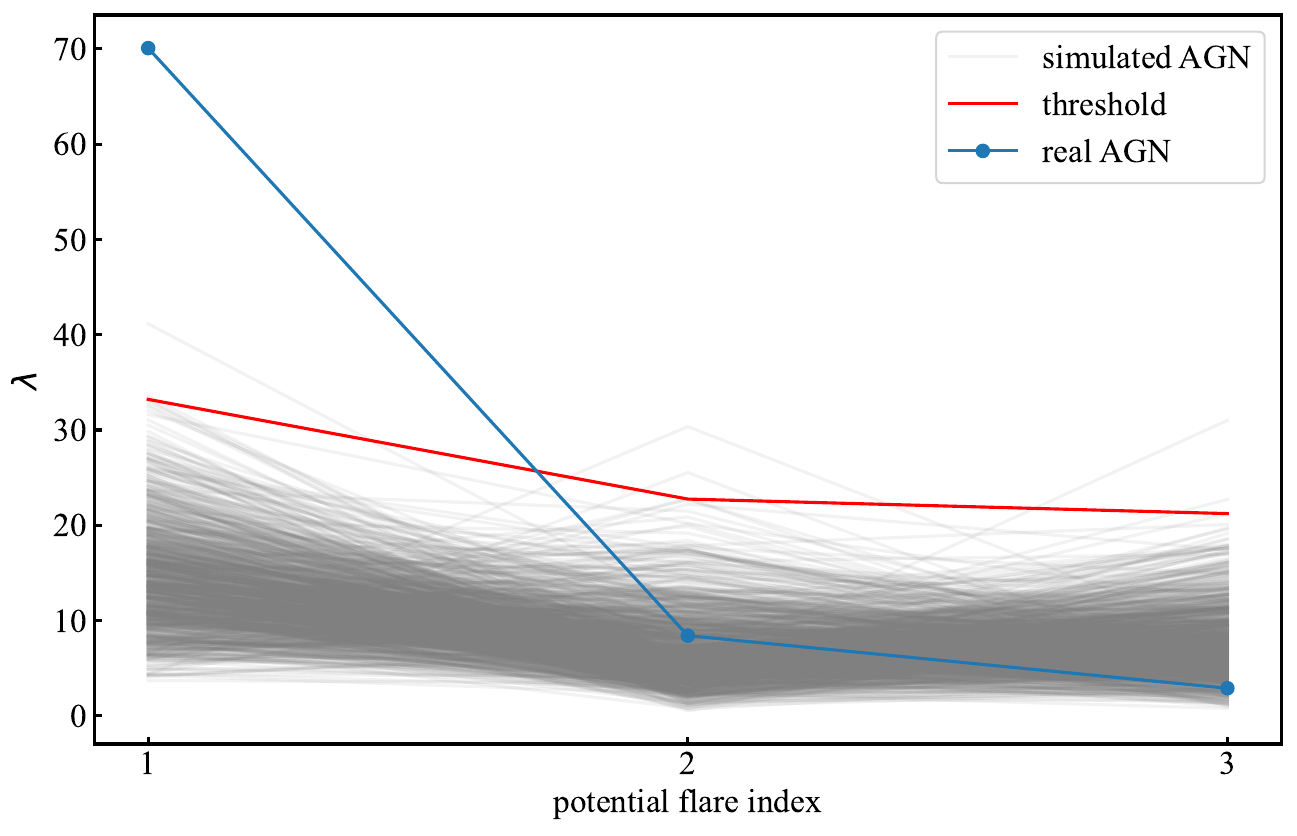}{0.5\textwidth}{(d)}
  }
  \caption{The detection processing flow for a real AGN light curve. Panel (a) shows the light curve of a real AGN (blue), with the red curve representing the fit obtained using Bayesian block representation. The green shaded regions indicate the time intervals where potential flares are detected. Panel (b) displays the simulated light curves, while panel (c) presents the GP hyperparameters for all the light curves. Panel (d) shows the $\lambda$ values computed for all light curves in the potential flare regions. One point corresponding to the real AGN exceeds the threshold derived from simulations, which corresponds to a flare significance of $p_{\mathrm{flare}} > 0.998$, indicating that this event is likely a genuine flare.}
  \label{fig:AGNExample}
\end{figure*}

Figure \ref{fig:AGNExample} illustrates the processing flow for a real AGN light curve, using the source J124942.29+344929.0 (RA=192.4262$^{\circ}$, Dec=34.8247$^{\circ}$) as an example. This AGN has a potential flare with $p_{\mathrm{flare}} = 1.0$ and is ultimately retained in our final high-confidence flare sample. Gaussian Process modeling and related computations are performed using the Python package \texttt{celerite}, which is optimized for fast and scalable GP analysis on one-dimensional data \citep{foreman-mackeyFastScalableGaussian2017}. Following this procedure, we identify 28,504 AGN flare events, which are summarized in the AGN Flare Coarse Catalog - Version 1.0 (AGNFCC-V1.0).

\subsection{Refinement and Selection of High-Confidence Flares}

While the initial flares identified using Gaussian Process modeling provide a valuable starting point, this coarse selection may still include false positives arising from irregular sampling or photometric noise. Moreover, given the large number of AGNs analyzed, some statistically significant events are expected to arised purely by chance, which is a manifestation of the look-elsewhere effect \citep{vaughanFalsePeriodicitiesQuasar2016}. To enhance the reliability of the flare sample, we apply a series of additional selection criteria designed to filter out ambiguous or low-significance events.

We remove light curves that exhibit observational gaps longer than 500 days in either the g and r bands first, as such gaps may hinder the reliable identification of flares.

To ensure that a flare represents a significant and localized enhancement in flux, we require that the peak flux in both bands exceed the global median by more than 3$\sigma$. Meanwhile, the remaining non-flare epochs must remain within this threshold, with at most one exception permitted per band. This condition is designed to isolate events that stand out clearly against the quiescent background, rather than being part of broader or irregular variability.

To assess whether a flare is statistically favored over baseline variability, we refit each light curve using a Gaussian Process model that incorporates a flare-shaped mean function, characterized by a Gaussian rise followed by an exponential decay:
\begin{equation}
  \label{eq:flare}
  y(t) = \left\{
    \begin{aligned}
      &r_{0} + A\exp\left(-\frac{(t-t_{0})^{2}}{2t_{g}^{2}}\right), t \leq t_{0} \\
      &r_{0} + A\exp\left(-\frac{(t-t_{0})}{t_{e}}\right), t \ge t_{0},
    \end{aligned}
    \right.
\end{equation}
where $r_{0}$ denotes the baseline flux level, $A$ is the flare amplitude, $t_{0}$ is the time of the flare peak, and $t_{g}$ and $t_{e}$ represent the characteristic timescales of the rising and decaying phases, respectively. This model is compared against a baseline GP with a constant mean function. We compute the Bayesian Information Criterion (BIC) for both models and retain flares only if the flare model yields a BIC improvement greater than 10, indicating strong evidence for the presence of a distinct flare.

Each flare is required to exhibit a well-sampled temporal profile. Specifically, the interval [$t_{0}-t_{g},t_{0}+t_{e}$], which defines the expected duration of the flare, must be fully covered by observations and contain more than $(t_{g}+t_{e})/20$ data points, corresponding to an average cadence of at least one observation every five days. Additionally, the flare amplitude must be significant, with the ratio $A/r_{0} > 0.2$, ensuring that the flare stands out clearly above the baseline flux level.

In addition, we exclude sources that show signs of blending or positional confusion arising from the 3$^{\prime\prime}$ cross-matching radius used in catalog construction. In particular, if multiple ZTF sources are associated with a single AGN and exhibit inconsistent separations, such as some within 1$^{\prime\prime}$ and others beyond 1.5$^{\prime\prime}$, we treat this as evidence of source contamination. In such cases, we extract only the light curve from sources within 1$^{\prime\prime}$ and recompute the flare significance following the procedure described in Section \ref{sec:flare_detection}.

All of the above selection criteria are summarized in Table \ref{tab:SelectionCriteria}, which also lists the number of flares removed by each individual condition. To better assess the relative impact and independence of each criterion, we also report the number of sources that are uniquely removed. These statistics provide insight into the discriminating power and redundancy among different filter rules. Representative examples of excluded flares are shown in Appendix \ref{appendix:excluded_lcs}, illustrating typical reasons for rejection.

\begin{deluxetable*}{cccc}[htb!]
  \tablecaption{Summary of Flare Selection Criteria}
  \tablehead{
    \colhead{number} & \colhead{Criterion Description}& \colhead{Removed} & \colhead{Uniquely Removed}
  }
  \label{tab:SelectionCriteria}
  \startdata
  \hypertarget{cri:1}{1} & Observational gap $>$ 500 days & 4228 & 56 \\
  \hypertarget{cri:2}{2}& Peak flux does not exceed 3$\sigma$ & 3823 & 118\\
  \hypertarget{cri:3}{3} & More than one non-flare point exceeds $3\sigma$ & 9947 & 1794 \\
  \hypertarget{cri:4}{4} & $\Delta \mathrm{BIC} < 10$ between flare and flat model & 16794 & 1583 \\
  \hypertarget{cri:5}{5} & Flare duration ($t_{0}-t_{g},t_{0} + t_{e}$) not fully covered & 10504 & 760 \\
  \hypertarget{cri:6}{6} & Too few data points within the flare interval & 16694 & 1096\\
  \hypertarget{cri:7}{7} & Flare amplitude $A/r_{0} \leq 0.2$ & 3502 & 147 \\
  \hypertarget{cri:8}{8} & Source confusion within $3^{\prime\prime}$ match radius& 974 & 26 \\
  \enddata
  \tablecomments{The \textit{Remove} column shows the number of flares that fail each criterion, while the \textit{Uniquely Removed} counts those excluded solely by that condition. Since some sources violate multiple criteria, the total number of removed flares is not equal to the number of sources in AGNFCC.}
\end{deluxetable*}

Based on the selection criteria described above, we have constructed a new catalog, referred to as the AGN Flare Refined Catalog - Version 1.0 (AGNFRC-V1.0), which contains 1,984 flares. These flares have passed the initial set of filters, ensuring that they exhibit significant variability and meet the physical and observational requirements for AGN flare events.

However, it is worth noting that the strict selection criteria are applied under the assumption that each AGN hosts at most one significant flare. In rare cases where an AGN exhibits multiple flares, the source may be easily excluded by some criteria. These multi-flare systems remain included in the broader AGNFCC and can be revisited using more flexible selection strategies. To facilitate this, the individual selection flags and related properties for each criterion are provided in the catalog, as detailed in Table \ref{tab:CatalogColumns}, allowing users to define their filtering schemes based on specific scientific goals.

\section{Catalog Demographics \label{sec:catalog}}
All columns of these two catalogs are presented in Table \ref{tab:CatalogColumns}. The catalogs themselves can be downloaded at \url{https://github.com/Lyle0831/AGN-Flares}.

\begin{deluxetable*}{lll}[htb!]
  \tablecaption{Columns of the catalogs.}
  \tablehead{
    \colhead{Property} & \colhead{Unit} & \colhead{Description}
  }
  \label{tab:CatalogColumns}
  \startdata
  AGN\_name & & Name of the host AGN. \\
  RA & deg & Right ascension (J2000).\\
  DEC & deg & Declination (J2000).\\
  z & deg & Redshift.\\
  source & & Source catalog: `M' for Milliquas, `D' for DESI, `L' for LAMOST, `Q' for Quaia, and `W' for WISE.  \\
  type & & Source type, including `Blazar', `SN', and `TDE' as described in Section \ref{sec:SourceClassification}\\
  ALeRCE & & Classification result from the ALeRCE as described in Section \ref{sec:SourceClassification}.\\
  \added{prob\_diff\_top2} & & \added{Difference between the highest and the second-highest class probabilities from the ALeRCE.}\\
  \added{alert\_offset} & \added{arcsec} & \added{Angular separation between the ZTF alert position (from ALeRCE) and the AGN coordinates.} \\
  t0 & MJD & $t_{0}$, peak time of the flare.\\
  te & day & $t_{e}$, exponential decay timescale of the flare.\\
  tg & day & $t_{g}$, gaussian rise timescale of the flare.\\
  r0\_g & $\mu \mathrm{Jy}$ & $r_{0g}$, baseline flux density in g band.\\
  r0\_r & $\mu \mathrm{Jy}$ & $r_{0r}$, baseline flux density in r band.\\
  A\_g & $\mu \mathrm{Jy}$ & $A_{g}$, flare amplitude in g band.\\
  A\_r & $\mu \mathrm{Jy}$ & $A_{r}$, flare amplitude in r band.\\
  max\_gap & day & Maximum observational gap in light curve. \\
  peak\_sigma\_g & & Significance level of the peak flux in g band.\\
  peak\_sigma\_r & & Significance level of the peak flux in r band.\\
  excess\_num\_g & & Number of non-flare points exceeding 3$\sigma$ in g band.\\
  excess\_num\_r & & Number of non-flare points exceeding 3$\sigma$ in r band.\\
  delta\_bic & & Difference between BIC for the constant model and the flare model. \\
  flare\_point\_g & & Number of points within the flare interval $(t_{0}-t_{g},t_{0}+t_{e})$ in g band.\\
  flare\_point\_r & & Number of points within the flare interval $(t_{0}-t_{g},t_{0}+t_{e})$ in r band.\\
  flag\_1 & & Boolean flag indicating whether max\_gap $>$ 500 days (criterion \hyperlink{cri:1}{1}).\\
  flag\_2 & & Boolean flag indicating whether min(peak\_sigma\_g, peak\_sigma\_r) $<$ 3 (criterion \hyperlink{cri:2}{2}).\\
  flag\_3 & & Boolean flag indicating whether excess\_num\_g + excess\_num\_r $>$ 2 (criterion \hyperlink{cri:3}{3}).\\
  flag\_4 & & Boolean flag indicating whether delta\_bic $<$ 10 (criterion \hyperlink{cri:4}{4})\\
  flag\_5 & & Boolean flag indicating whether t0 +te $>$ max(mjd) or t0 - tg $<$ min(mjd) (criterion \hyperlink{cri:5}{5}).\\
  flag\_6 & & Boolean flag indicating whether min(flare\_point\_g, flare\_point\_r) $<$ (tg+te)/20 (criterion \hyperlink{cri:6}{6}).\\
  flag\_7 & & Boolean flag indicating whether min(A\_g/r0\_g, A\_r/r0\_r) $<$ 0.2 (criterion \hyperlink{cri:7}{7})\\
  flag\_8 & & Boolean flag indicating whether this AGN is confused with other sources (criterion \hyperlink{cri:8}{8}).\\
  \enddata
\end{deluxetable*}

\subsection{Overall Characteristics}
The number of sources in each catalog is shown in Table \ref{tab:SourceNumber}. Starting with approximately 2 million AGN light curves with sufficient ZTF observations, we identify 28,504 AGN flares in the AGNFCC, accounting for 1.5\% of the total. The AGNFRC contains 1,984 flares, representing 0.1\% of the total.

\begin{figure}[htb!]
  \includegraphics[width=0.45\textwidth]{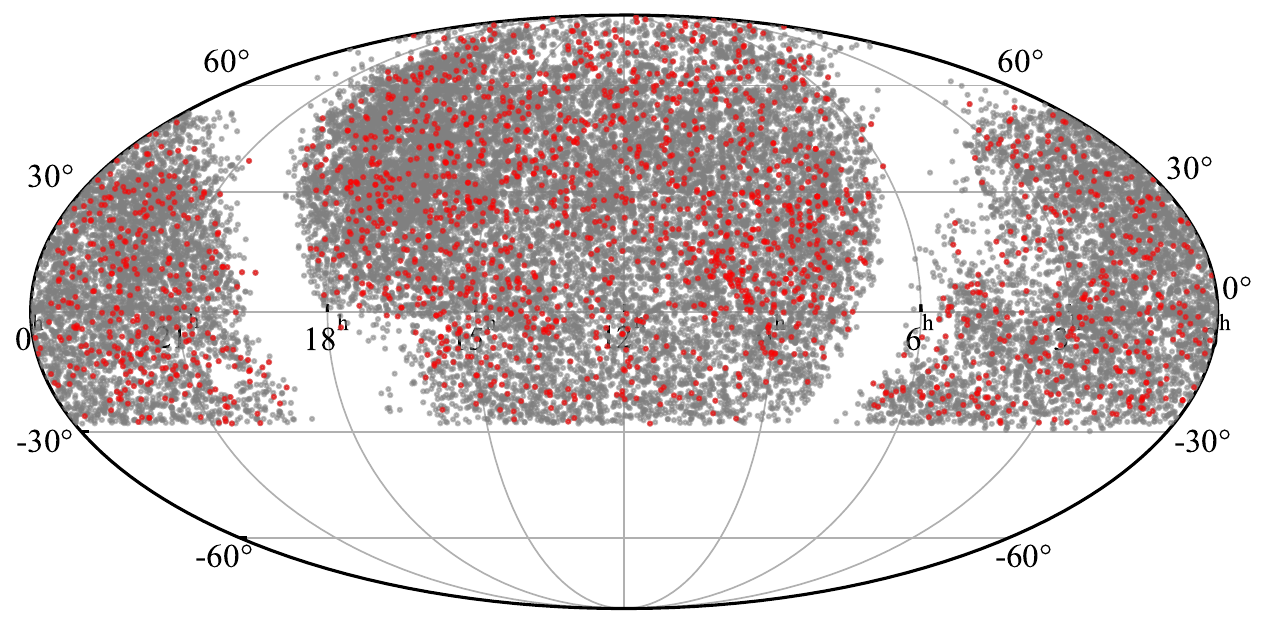}
  \caption{Sky distribution of AGN flares. Gray points represent flares from the AGNFCC, while red points correspond to flares from AGNFRC.}
  \label{fig:AGNFlareLocation}
\end{figure}

The spatial distribution of the AGN flares in the two catalogs is shown in Figure \ref{fig:AGNFlareLocation}. As expected, the flares are distributed relatively uniformly at declinations above $-30^{\circ}$ and outside the Galactic plane, consistent with the underlying AGN distribution in the AGN catalogs and the sky coverage of the ZTF survey.

\begin{figure}[htb!]
  \centering
  \includegraphics[width=0.45\textwidth]{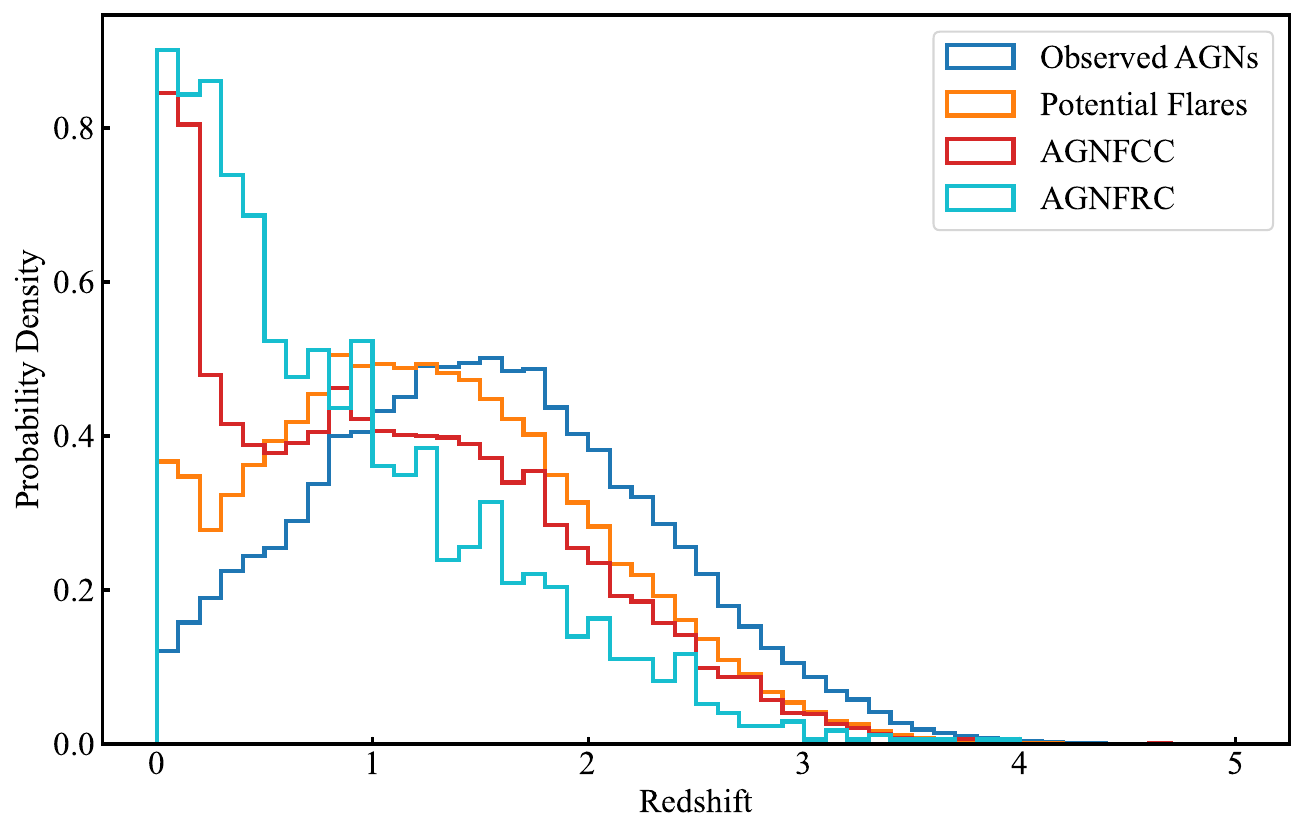}
  \caption{Redshift distribution of different samples.}
  \label{fig:AGNFlareRedshift}
\end{figure}

We present the redshift distributions for four samples in Figure \ref{fig:AGNFlareRedshift}, including all AGNs with more than 30 observations per band, potential flares identified via Bayesian Block representation, and the AGN flares in the AGNFCC and AGNFRC. There is a clear trend that samples with increasingly reliable flares show progressively stronger low-redshift concentration. This suggests that AGN flares are more readily detected in nearby sources. At higher redshifts, only intrinsic bright AGNs can be observed, and in these luminous systems, flares of similar absolute brightness produce weaker contrast relative to the AGN baseline flux, making them harder to detect. \added{Furthermore, the excess at low redshift may be partly due to contamination from supernovae, which produce transient signals offset from the galactic nucleus. To account for this possibility, we have included the parameter \texttt{alert\_offset} in the catalog, representing the angular distance between the transient and the AGN position, thereby assisting users in identifying and filtering out potential off-nucleus supernovae contamination.}

\begin{figure*}[htb!]
  \centering
  \gridline{\fig{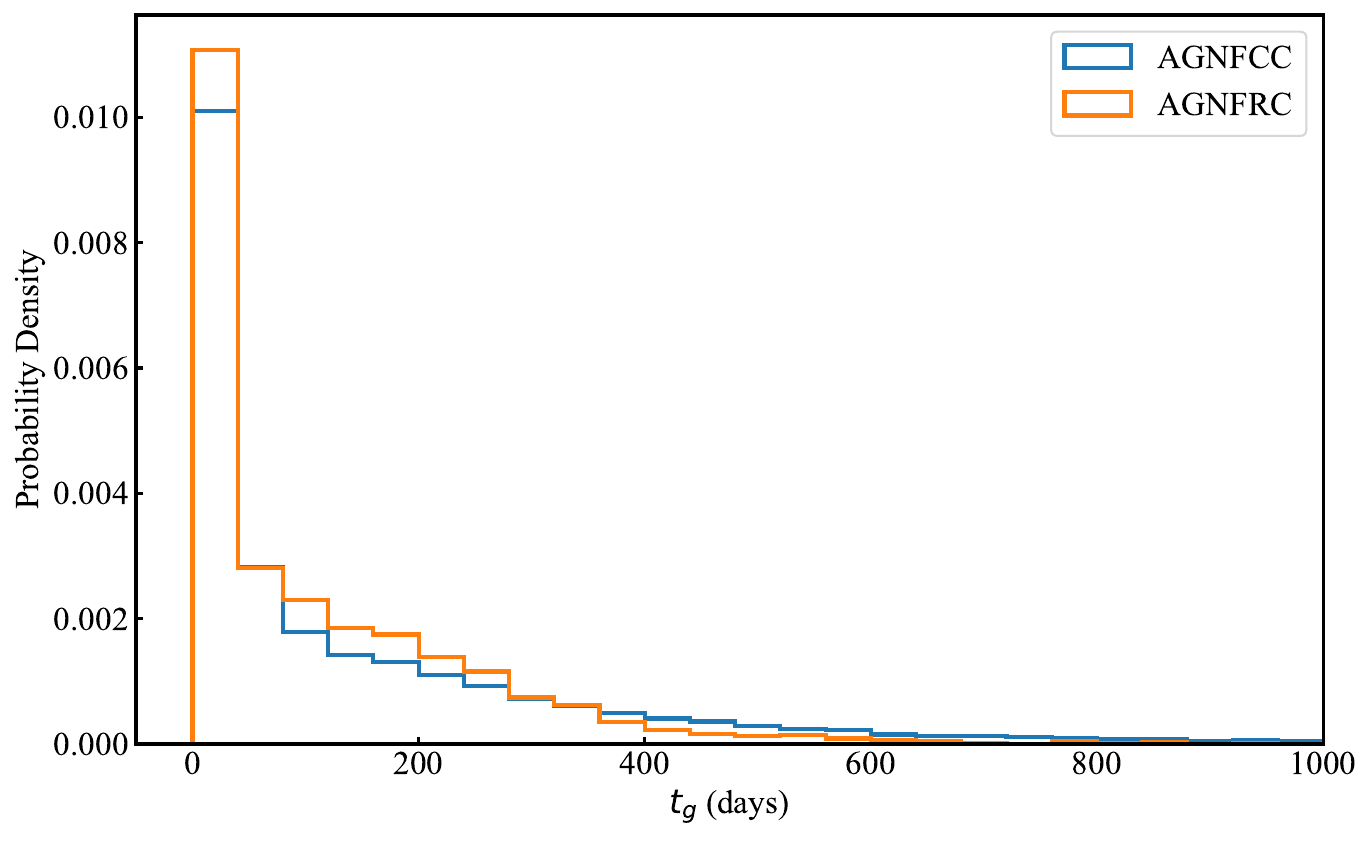}{0.5\textwidth}{}
            \fig{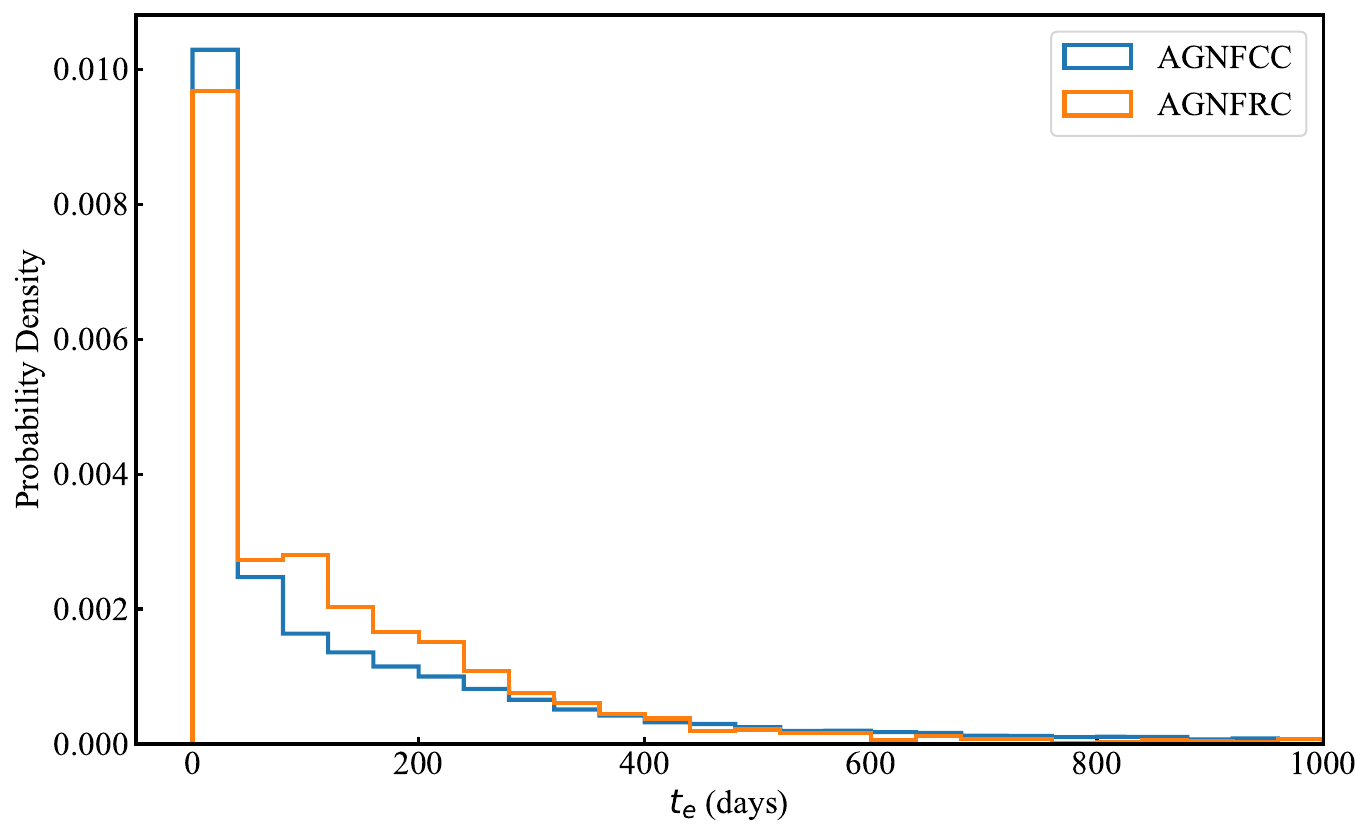}{0.5\textwidth}{}
          }
  \caption{Distribution of rise timescale $t_{g}$ (left) and decay timescale $t_{e}$ (right) for two AGN flare catalogs.}
  \label{fig:flareTimeScale}
\end{figure*}

Figure \ref{fig:flareTimeScale} presents the distributions of the flare rise and decay timescales, denoted as $t_{g}$ and $t_{e}$ in Equation (\ref{eq:flare}), respectively. The timescales span a broad range, from just a few days to several hundred days, reflecting the diverse nature of AGN flares. The majority of flares in both AGNFCC and AGNFRC exhibit rise and decay timescales shorter than 50 days. However, the very shortest timescales should be interpreted with caution, as our 3-day binning may introduce uncertainties for such rapid flares. \added{We also note that the derived timescales are influenced by the annual gaps in the ZTF light curves. These gaps can truncate the observable segment of long-duration flares, potentially biasing the measured timescales, while short-duration flares occurring entirely within the gaps may be missed. Therefore, the presented distributions reflect only the detectable range under the current observational limitations.}

\begin{figure}[htb!]
  \centering
  \includegraphics[width=0.45\textwidth]{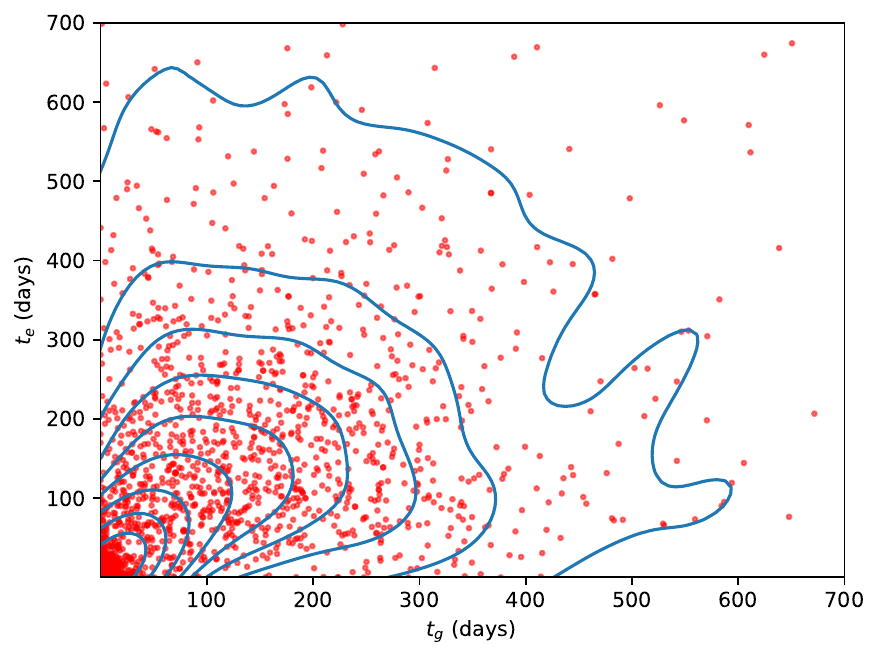}
  \caption{Joint distribution of rise timescale $t_{g}$ and decay timescale $t_{e}$ for flares in the AGNFRC. Red points represent individual flares, while blue contours indicate the kernel density estimate of their distribution.}
  \label{fig:tg-te}
\end{figure}

\begin{figure*}[htb!]
  \centering
  \gridline{\fig{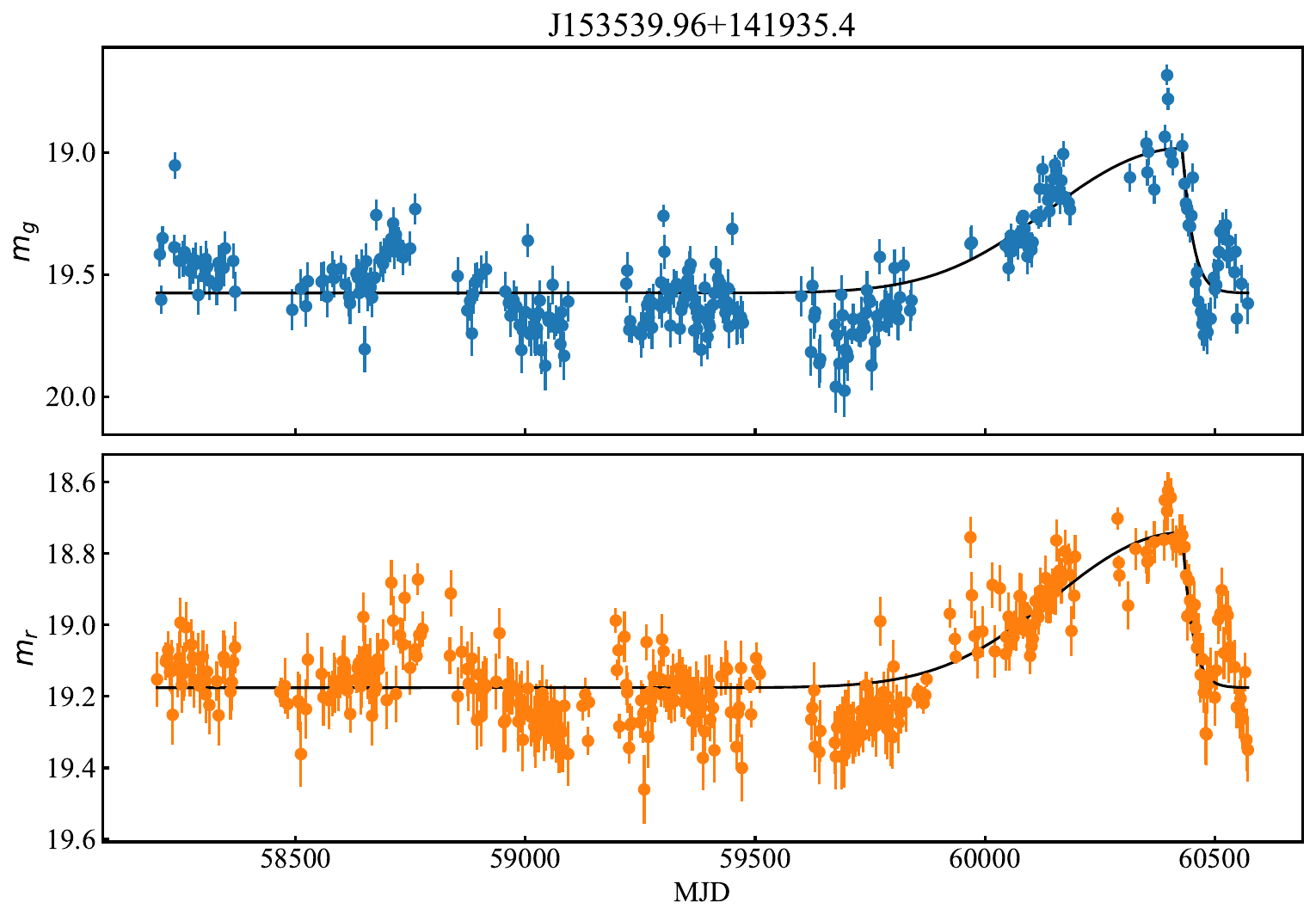}{0.5\textwidth}{}
    \fig{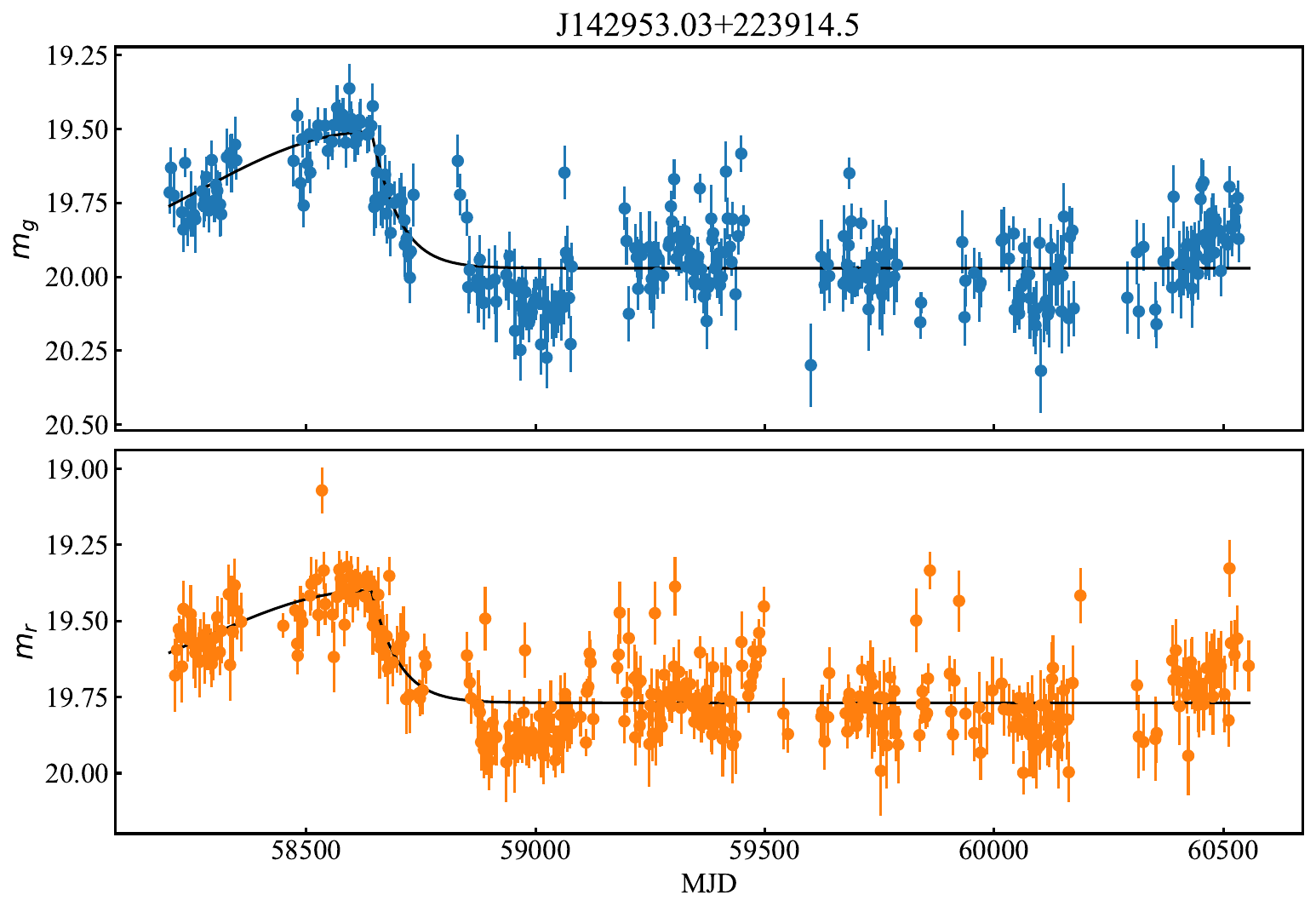}{0.5\textwidth}{}
  }
  \caption{Light curves of AGN flares that have a slow-rise/fast-decay profile. The black lines indicate the best-fitting profiles using Equation (\ref{eq:flare}).}
  \label{fig:Lightcurve-slowrise}
\end{figure*}

Additionally, the flare profiles also show considerable variety, including fast-rise/slow-decay and slow-rise/fast-decay patterns (see Figure \ref{fig:tg-te}). The latter type appears relatively rare, and two representative light curves are presented in Figure \ref{fig:Lightcurve-slowrise}. Such diversity may be associated with different physical processes, potentially linked to AMSs located in different regions of the AGN disk \citep{liuAccretionmodifiedStarsAccretion2024}.

\subsection{Host AGN Types}
\begin{figure}[htb!]
  \includegraphics[width=0.45\textwidth]{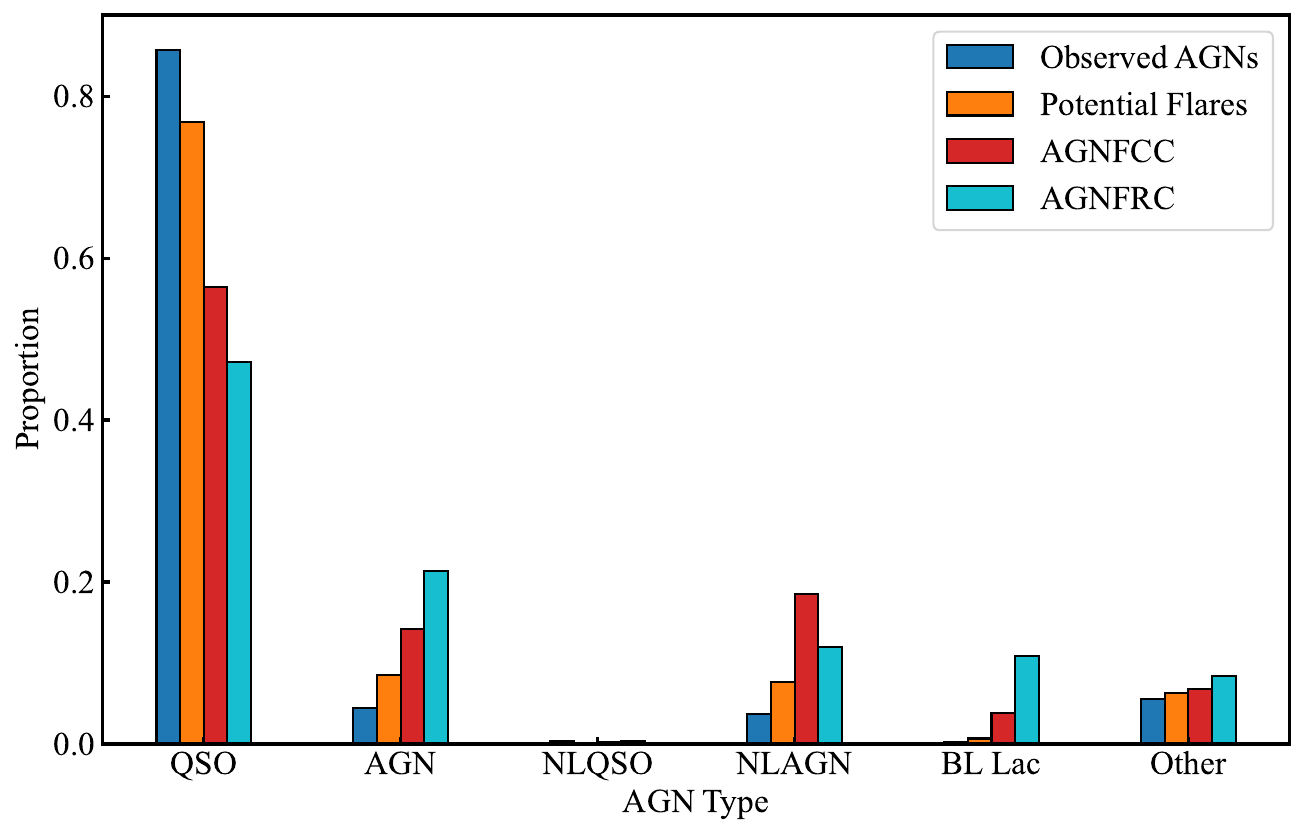}
  \caption{AGN type proportions in Million Quasars Catalog.}
  \label{fig:AGNType}
\end{figure}

Based on the classifications provided in the Million Quasars Catalog, we present the fractional distribution of AGN types across different samples in Figure \ref{fig:AGNType}. Quasars (QSOs) dominate the overall observed AGN population, accounting for over 80\% of the sources. However, their relative proportion decreases steadily from the general AGN sample to the flare catalogs. In contrast, sources classified as ``AGN'' become increasingly prominent in the flare samples. This trend suggests that non-quasar AGNs are more likely to exhibit flaring event detectable to our methods, which is expected as quasars are so luminous that normal flares become difficult to distinguish against their bright background.

We also observe notable changes in the relative fractions of narrow-line AGNs (NLAGNs) and BL Lac objects. Similar to the trend seen in the sources classified as ``AGN'', the fraction of NLAGNs increases significantly in the flare samples, suggesting that flares are more detectable in low-luminosity systems. However, their proportion decreases from AGNFCC to AGNFRC, indicating that many of the AGN flares initially identified did not pass the stricter criteria for flare reliability. A closer inspection reveals that 84\% of the NLAGNs in AGNFCC are excluded from AGNFRC due to the presence of multiple points exceeding 3$\sigma$ in non-flare regions, compared to only 35\% for the full AGNFCC sample. This suggests that NLAGNs are more prone to exhibiting irregular or noisy variability patterns that mimic flare-like features, resulting in a higher false positive rate during the initial detection. Nonetheless, the occurrence of prominent flares in NLAGNs is relatively rare because the accretion disks of these systems are generally considered to be obscured by a dusty torus \citep{urryUnifiedSchemesRadioLoud1995}. However, these flares can be observed in so-called ``true'' type 2 AGNs~\citep{Tran2001,Shi2010,Barth2014}, in which the lack of broad-line regions (BLRs) is intrinsic, or turn-off AGNs, in which the BLRs have disappeared very recently~\citep{Lamassa2015,Runnoe2016}.
In this scenario, a sudden increase in the accretion rate causes dramatic luminosity enhancements and the emergence of BLRs, a phenomenon widely known as changing-look (turn-on) AGNs~\citep{Sheng2017,gezariIPTFDiscoveryRapid2017,zeltynTransientChanginglookActive2022}.

As for BL Lac objects, although they make up a small fraction overall, their proportion steadily increases and becomes more prominent in the refined sample. This trend is probably due to their strong and rapid variability, which makes them more likely to be identified as flares by our method.

\subsection{Source Classification\label{sec:SourceClassification}}

We begin by selecting blazars, known for their strong variability signals. By crossmatching the flares in the AGNFCC with sources from the Roma-BZCAT Multifrequency Catalogue of Blazars \footnote{\url{https://www.ssdc.asi.it/bzcat/}} \citep{massaro5thEditionRomaBZCAT2015} using a 3$^{\prime\prime}$ matching radius, and combining them with the blazars flagged in the Million Quasars Catalog, we identify a total of 830 blazars. These sources are expected to exhibit significant variability, making them a part of our catalog.

Next, we perform a crossmatch with a 3$^{\prime\prime}$ matching radius between the flares in the AGNFCC and the sources in the Transient Name Server \footnote{\url{https://www.wis-tns.org/}}  (TNS) database, focusing on supernovae (SNe) and tidal disruption events (TDEs). \added{We further check the temporal coincidence between the flares and the reported transients in the TNS database, and retain only those cases that are contemporaneous. After that, we find that 2 flares are associated with TDEs and 59 with SNe.} This suggests that a subset of the flares may be related to transient phenomena other than AGN variability, further emphasizing the diversity of potential sources contributing to the observed flaring activities.

We also use the Automatic Learning for the Rapid Classification of Events (ALeRCE) broker classifier \citep{sanchez-saezAlertClassificationALeRCE2021} of the ZTF alert stream\footnote{\url{https://alerce.online/}} to assign classifications to the AGN flares. \added{By crossmatching the flares with ZTF alerts within a 3$^{\prime\prime}$ matching radius, we obtain the corresponding probabilistic classifications. For each source, we adopt the class with the highest probability assigned by the ALeRCE classifier and calculate the difference between the two highest classification probabilities (\texttt{prob\_diff\_top2}) as an indicator of the classification confidence. Approximately 18\% of the sources have \texttt{prob\_diff\_top2} $<$ 0.1, suggesting relatively uncertain classifications. In practice, we mainly rely on the results from the Lc Classifier, and when those are unavailable, we adopt the classifications provided by the Stamp Classifier. Tohether, these two classifiers cover the majority of the ZTF alerts. However, it is crucial to note that these classifications are based solely on photometric data and thus have limited reliability. They are included in our catalog only as supplementary information for user reference.}

\begin{figure*}[htb!]
  \centering
  \gridline{\fig{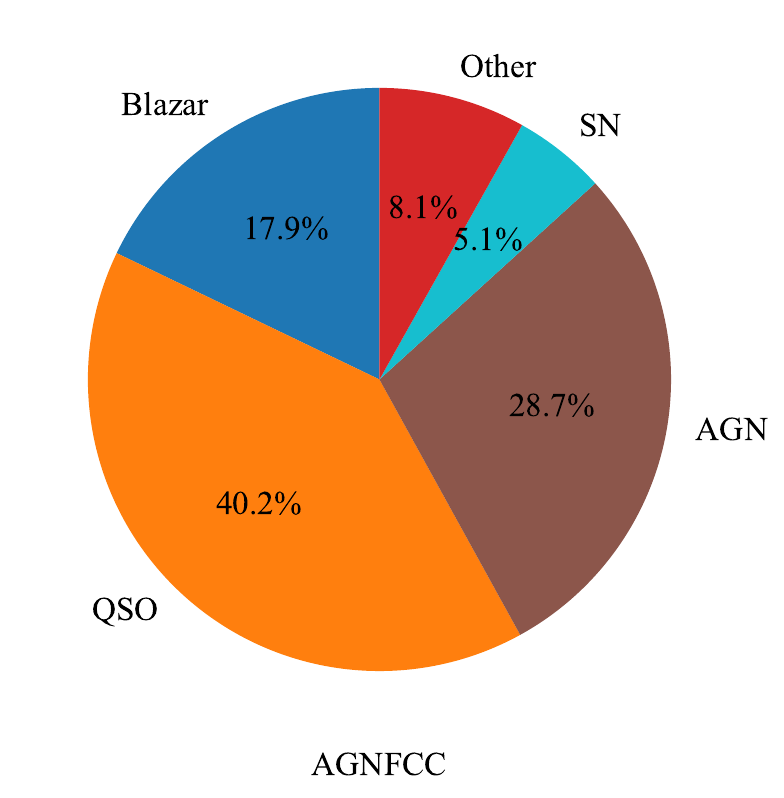}{0.45\textwidth}{}
    \fig{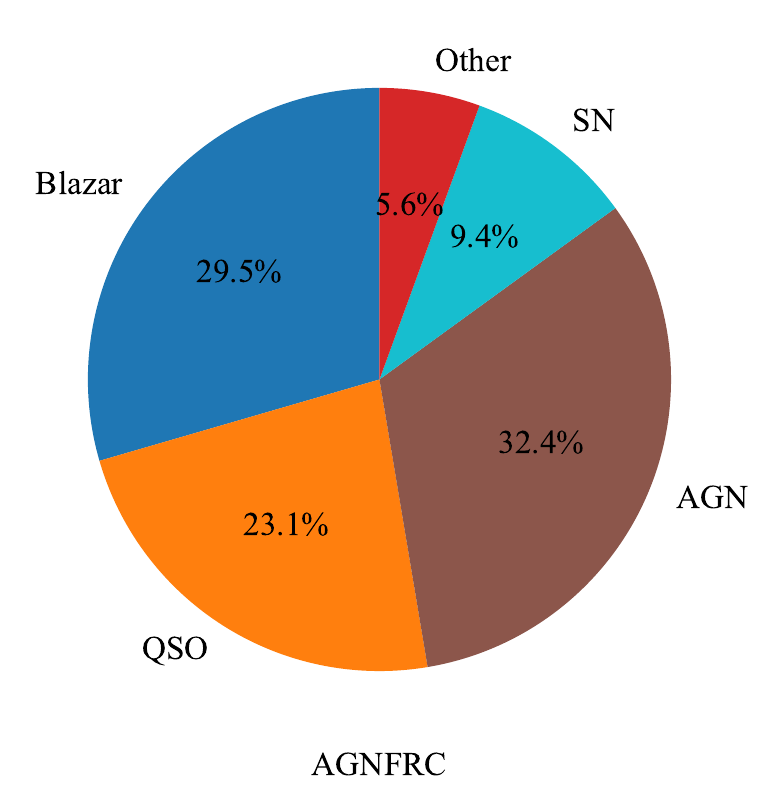}{0.45\textwidth}{}
  }
  \caption{Classification results of AGN flares in both AGNFCC (left) and AGNFRC (right) based on ZTF alerts.}
  \label{fig:ZTFAlert}
\end{figure*}

Since ALeRCE provides a large number of subclasses, we regroup them into five broad categories: SN, AGN, QSO, blazar, and other. In the AGNFCC, approximately 40\% of the flares can be matched with corresponding ZTF alert classifications, while this proportion increases to 70\% in the AGNFRC. This result demonstrates that the stricter selection criteria applied in the refined catalog are effective in identifying more reliable flares. Figure \ref{fig:ZTFAlert} shows the classification results of AGN flares in both catalogs. In the AGNFCC, the majority of flares are classified as QSOs, followed by AGNs and blazars. In the AGNFRC, the proportion of QSOs decreases, while the fractions of AGNs and blazars increase, mirroring the trend observed in Figure \ref{fig:AGNType}. As for SN, which represents a transient source rather than AGN-related variability, its proportion remains relatively small but increases slightly from AGNFCC to AGNFRC.

\section{discussion\label{sec:discussion}}
In this section, we explore several potential origins for the flares identified in our sample.

\subsection{Tidal Disruption Events}

When a star ventures too close to a black hole (BH), the extreme tidal field can overcome the star's self-gravity and tear it apart. The infall and accretion of the disrupted stellar material can give rise to a bright, short-lived flare known as a tidal disruption event \citep[TDE,][]{reesTidalDisruptionStars1988,gezariTidalDisruptionEvents2021}. Although TDEs have thus far predominantly been found in inactive galaxies, TDEs are also possible to occur within the dense, gas-rich environments of AGN accretion disks with an even higher rate~\citep{karasEnhancedActivityMassive2007,chanTidalDisruptionEvents2019,mckernanStarfallHeavyRain2022,ryuInplaneTidalDisruption2024,kaurElevatedRatesTidal2025}. These events may involve the disruption of main-sequence stars orbiting near the central SMBH \citep{ryuTidalDisruptionsMainsequence2020}, or occur around stellar-mass black holes embedded in AGN disks, leading to luminous signals \citep{yangTidalDisruptionStellarmass2022}.

\begin{figure*}[htb!]
  \centering
  \gridline{\fig{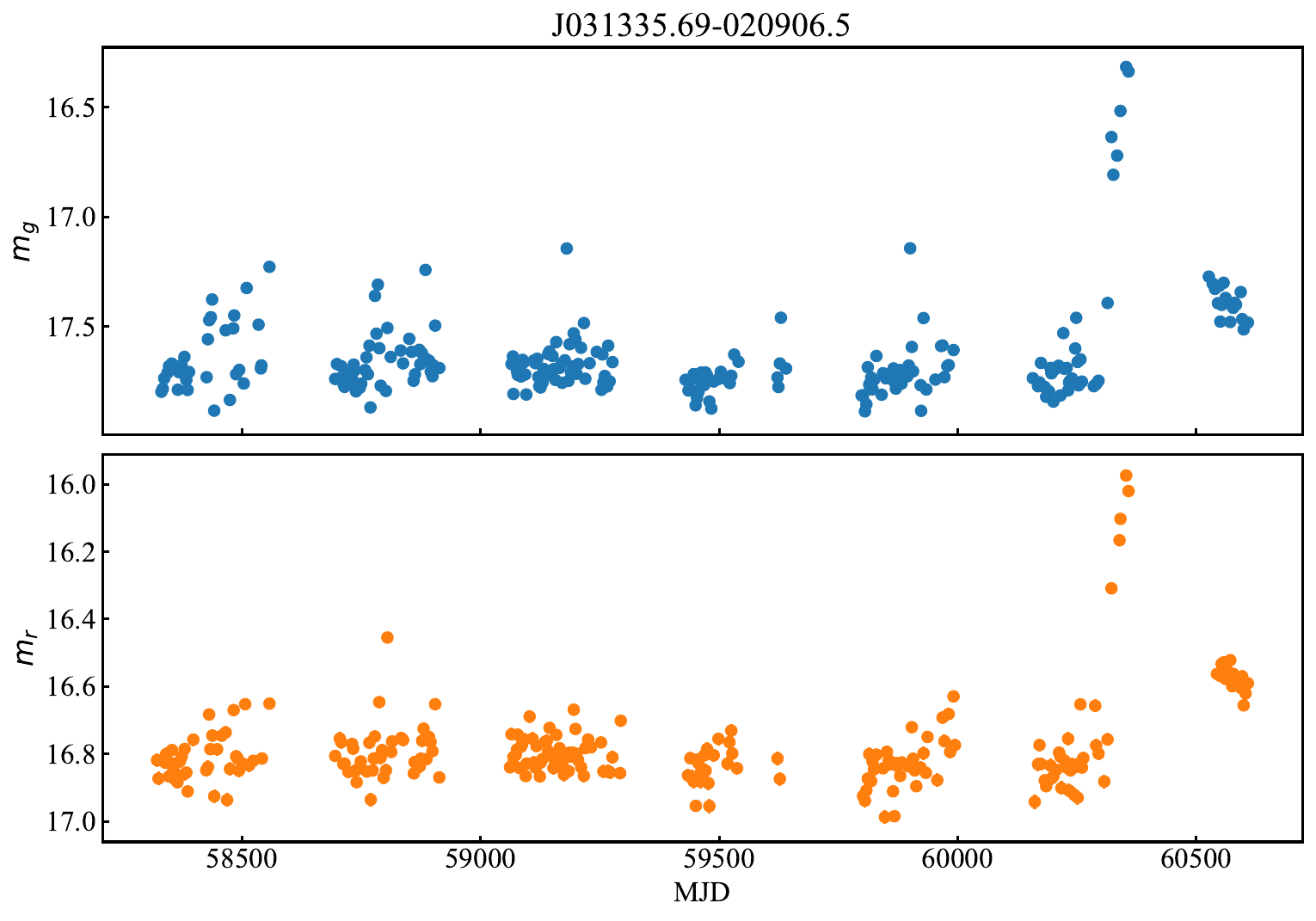}{0.5\textwidth}{}
    \fig{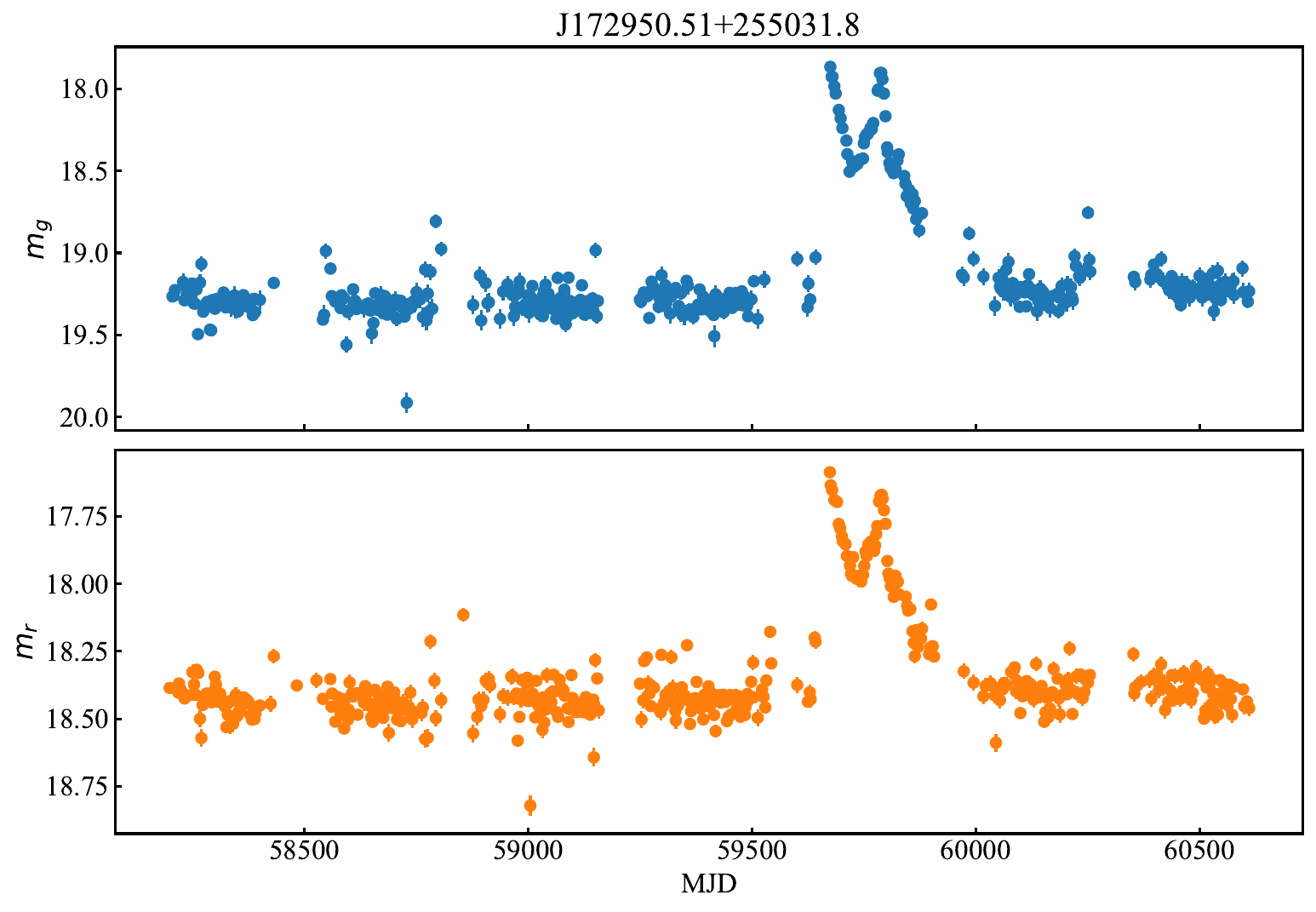}{0.5\textwidth}{}
  }
  \caption{Light curves of AT2020afhd (left) and AT2022exr (right), both classified as TDEs.}
  \label{fig:Lightcurve-TDE}
\end{figure*}

In the AGNFCC, we find two AGN flares that match with transients classified as TDEs in the TNS: AT2020afhd and AT2022exr. Both of them were classified as TDEs based on their persistent blue optical colors, UV emission, and supporting optical spectroscopic features \citep{hammersteinZTFTransientClassification2022, hammersteinClassificationAT2020afhdZTF20abwtifz2024}. AT2020afhd exhibits a rising timescale $t_{g}\sim 20$ days and a decay timescale $t_{e}\sim 120$ days. However, it is excluded from the AGNFRC due to insufficient data coverage during the decay phase, which failed to meet our criterion \hyperlink{cri:6}{6}. In contrast, AT2022exr shows similar timescales and is included in the AGNFRC. This source is particularly interesting because it exhibits repeating flare behavior, which remains challenging to fully explain with current models \citep{langisRepeatingFlaresXray2025}. The light curves of both sources are shown in Figure \ref{fig:Lightcurve-TDE}.

\subsection{Supernovae}

Supernovae (SNe) are violent explosions caused by core collapse of massive stars (core collapse supernovae, CCSNe) or thermonuclear explosions of white dwarfs (SNe Ia). SNe can occur spatially near the galactic center or within AGN accretion disks, where the environment could potentially enhance the rate of SNe \citep{grishinSupernovaExplosionsActive2021}. For example, the rate of CCSNe in AGN disks may increase due to more active star formations in these regions \citep{artymowiczStarTrappingMetallicity1993}. On the other hand, the rate of SNe Ia could also increase with a higher frequency of double white dwarf (WD) mergers \citep{mckernanBlackHoleNeutron2020}, or by accretion onto a single WD, leading to a thermonuclear explosion where the mass is close to the Chandrasekhar limit \citep{ostrikerViscousDragAccretion1983}. Both CCSNe and SNe Ia in AGN disks have been suggested as potentially observable events, with their signatures detectable in the AGN light curves \citep{liCorecollapseSupernovaExplosions2023, zhuThermonuclearExplosionsAccretioninduced2021}.

\begin{figure*}[htb!]
  \centering
  \gridline{\fig{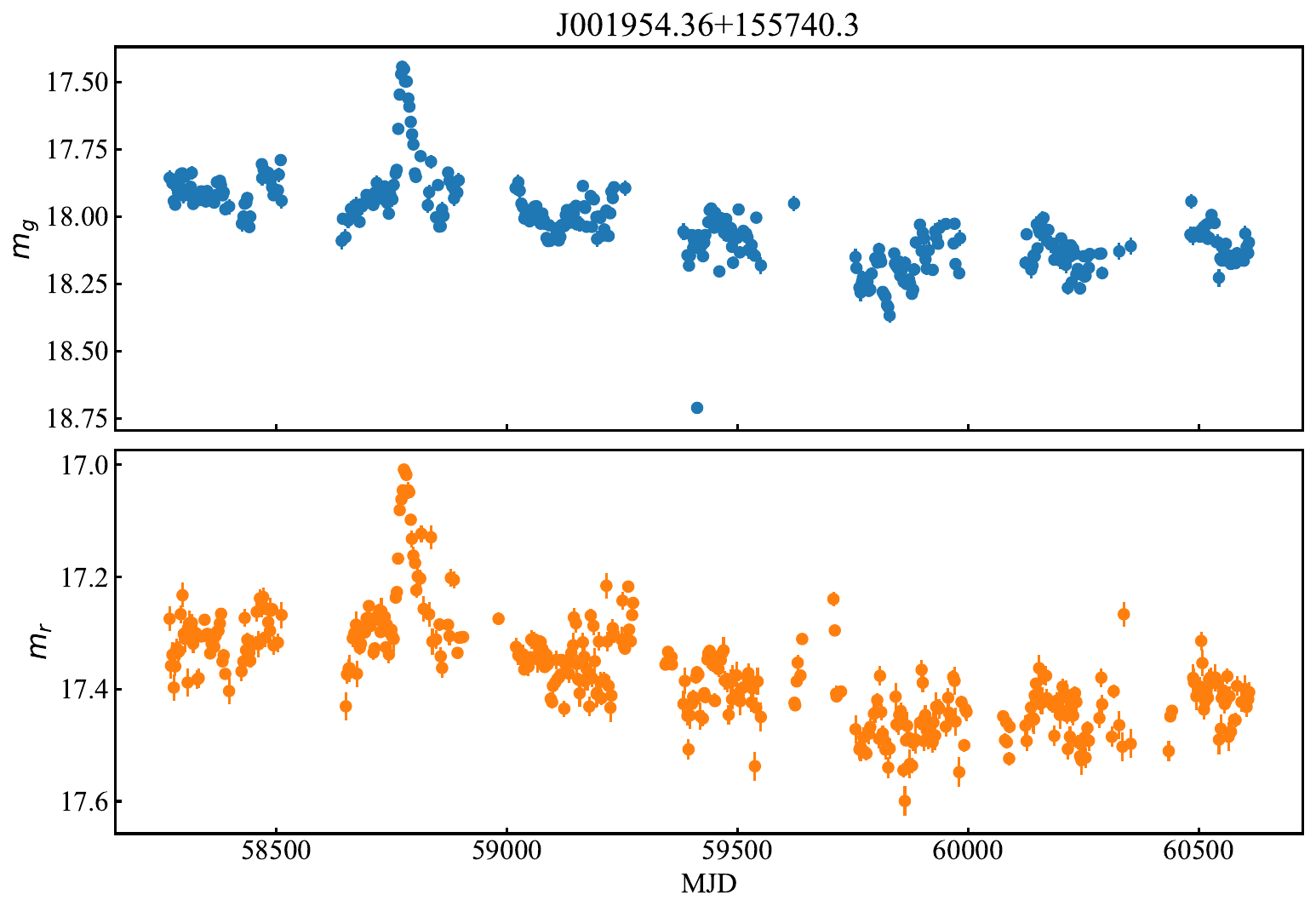}{0.5\textwidth}{(a)}
    \fig{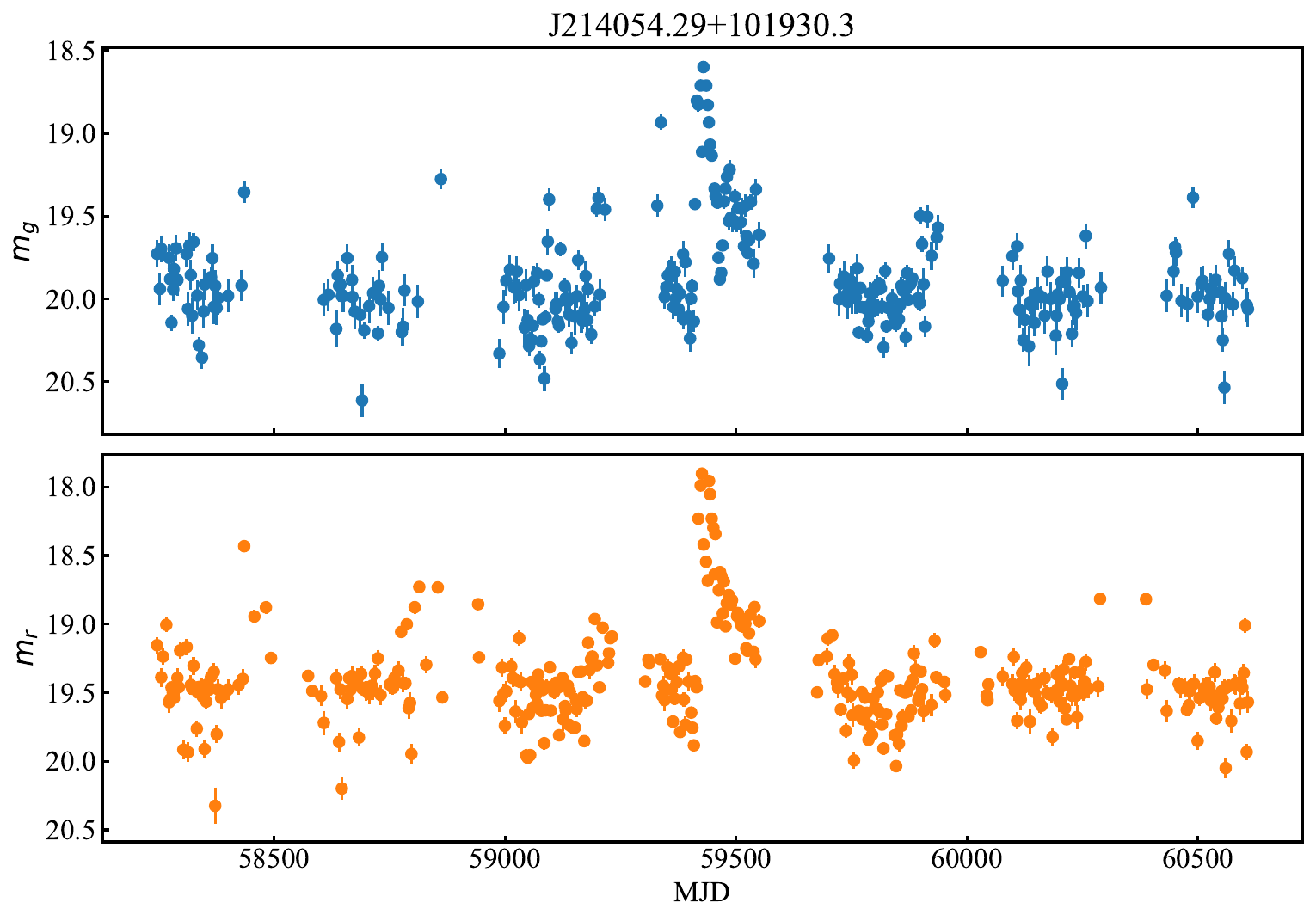}{0.5\textwidth}{(b)}
  }
  \gridline{\fig{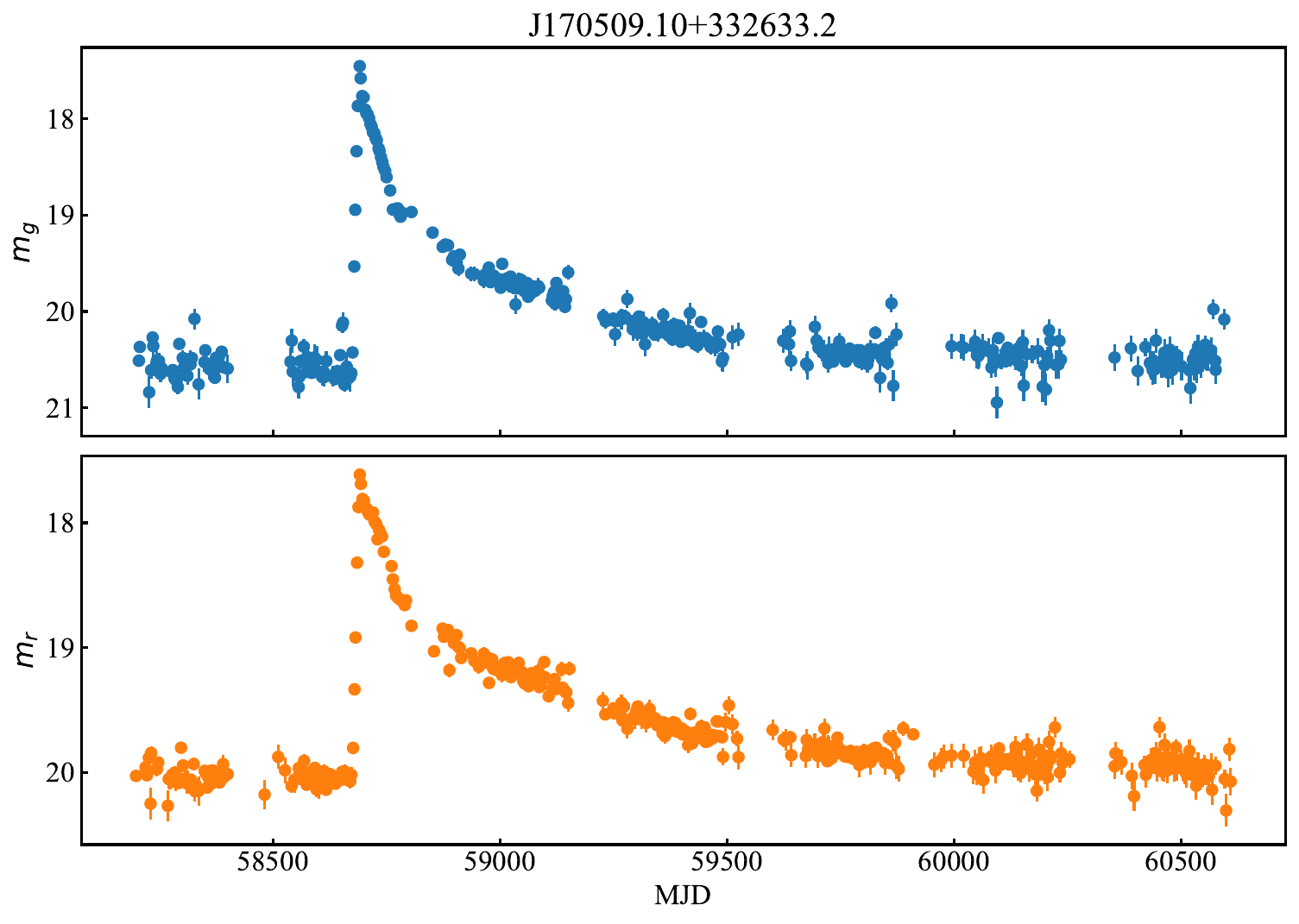}{0.5\textwidth}{(c)}
    \fig{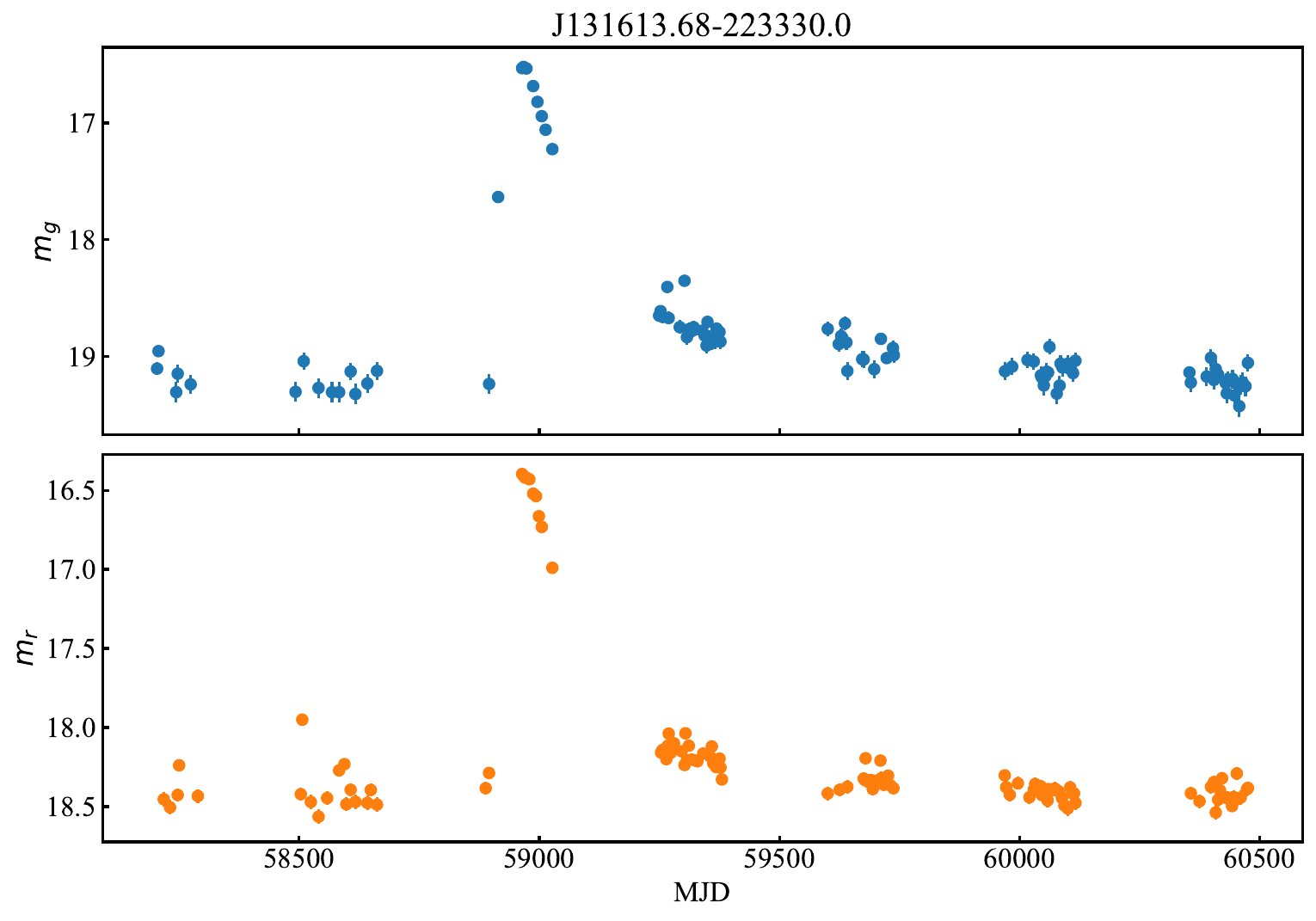}{0.5\textwidth}{(d)}
  }
  \caption{Light curves of AGN flares that matched with a known SN: (a) SN~2019ses (SN Ia), (b) SN~2021too (SN Ic), (c) SN~2019lkw (SN II), and (d) SN~2020edi (SLSN-II).}
  \label{fig:Lightcurve-SN}
\end{figure*}

After cross-matching AGNFCC sources with classification reports on the Transient Name Server (TNS), we find a total of 71 SNe spanning different classes in the AGNFCC sample. Most of the SNe are SNe Ia, which is likely due to higher intrinsic luminosities of SNe Ia than those of typical CCSNe, making them easier to identify even at galaxy centers. Figure \ref{fig:Lightcurve-SN} shows four representative light curves corresponding to different SN classes. In each case, the AGN shows typical variability outside the flare event, with the ``flare" feature representing a distinct and significantly brighter episode compared to the usual flux. However, observationally it is difficult to answer whether the SNe originate within AGN accretion disks or just spatially near galaxy centers.

\subsection{ Stellar mass Binary Black Hole Mergers}

Stellar mass binary black hole (BBH) mergers may occur within AGN accretion disks, which provide an environment that favors both the formation and retention of massive black holes. The deep gravitational potential of AGN disks makes it difficult for merger remnants to escape \citep{varmaEvidenceLargeRecoil2022}, enabling hierarchical mergers that gradually build up BH mass \citep{liMultimessengerHierarchicalTriple2025a}. In addition, the presence of dense gas facilitates the pairing and the orbital evolution of black holes through dynamical friction and gas torques \citep{yangHierarchicalBlackHole2019, mckernanMonteCarloSimulations2020}, making AGN disks efficient for producing high-mass BBH systems \citep{samsingAGNPotentialFactories2022}.

Observationally, there is evidence that a fraction of BBH mergers detected by the LIGO-Virgo-KAGRA (LVK) collaboration may originate from AGN by investigating the spatial correlation between gravitational-wave sky maps and the position of known AGNs \citep{zhuEvidenceFractionLIGO2025}. If BBH mergers occur in AGN disks, they may produce detectable electromagnetic (EM) counterparts due to interactions between the binary (or post-merger remnant) and the surrounding medium. Possible mechanisms include ram-pressure stripping of ambient gas \citep{mckernanRampressureStrippingKicked2019}, hyper-Eddington accreting leading to Bondi explosions \citep{wangAccretionmodifiedStarsAccretion2021}, or jet launching following recoil kicks \citep{chenElectromagneticCounterpartsPowered2024, tagawaShockCoolingBreakout2024,  rodriguez-ramirezOpticalUVFlares2025, maElectromagneticFlaresAssociated2025}.

\begin{figure*}
  \centering
  \gridline{\fig{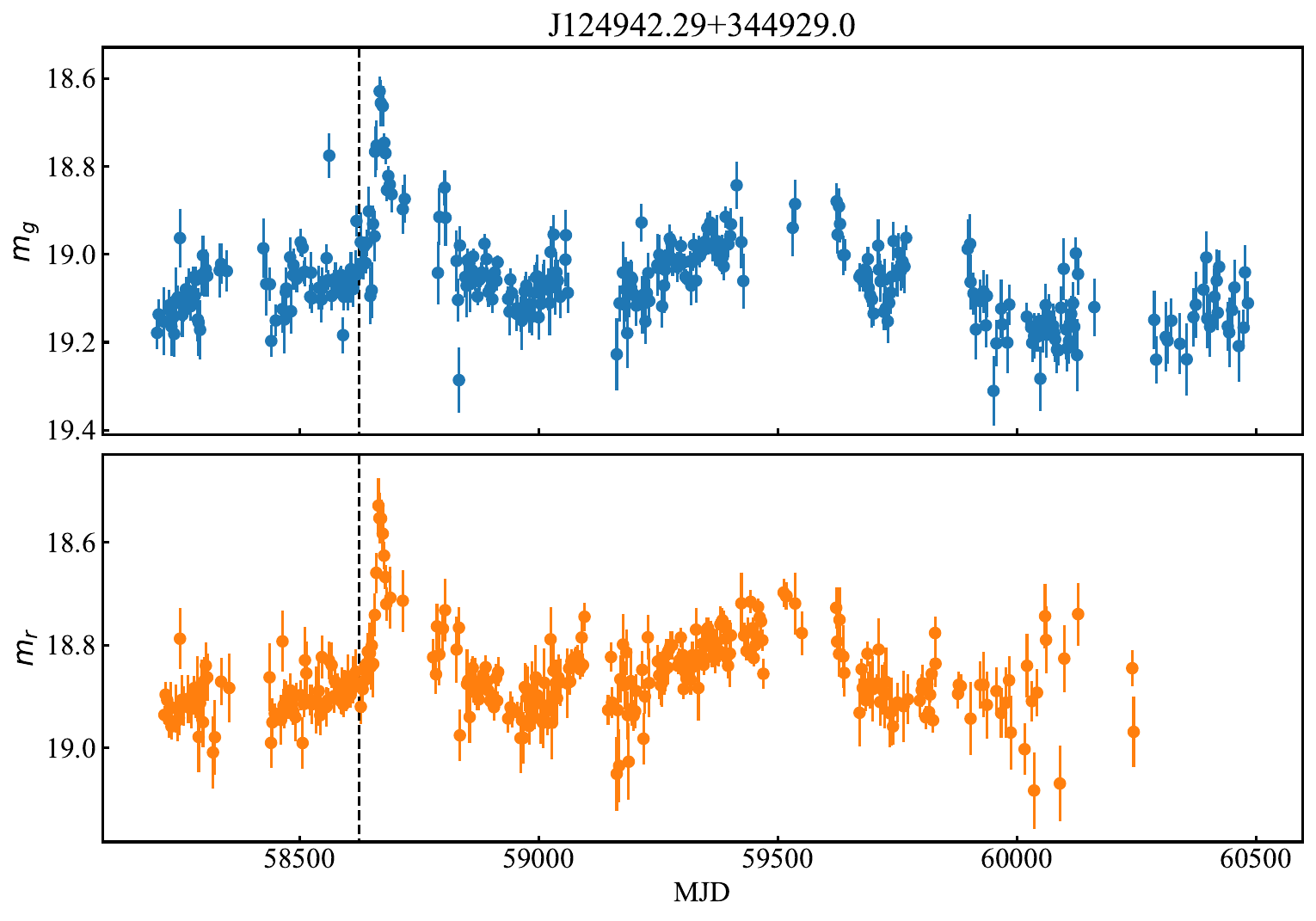}{0.5\textwidth}{}
    \fig{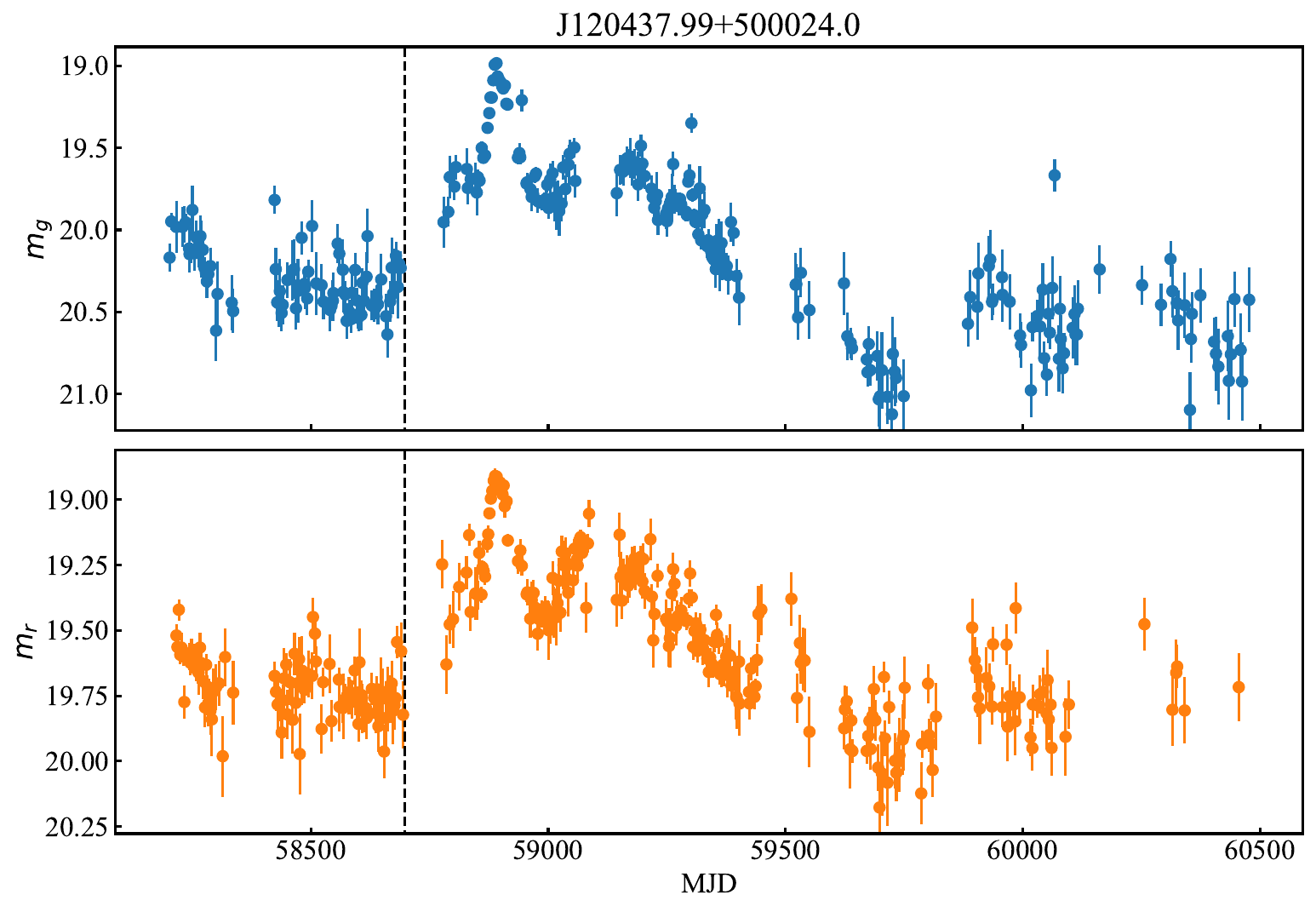}{0.5\textwidth}{}
  }
  \caption{Light curves of AGN flares that are potential EM counterparts to BBH mergers. The vertical dashed lines indicate the trigger times of the corresponding GW events. }
  \label{fig:Lightcurve-BBH}
\end{figure*}

\citet{grahamLightDarkSearching2023} reported seven AGN flares that may be associated with BBH mergers based on their temporal and spatial coincidence with GW events. An updated analysis by \citet{heTracingLightIdentification2025} using additional data found that only two of these flares maintain a strong correlation with GW events. Both of them are included in our AGNFRC, and their light curves are shown in Figure \ref{fig:Lightcurve-BBH}.

\subsection{Microlensing}

Microlensing caused by the close passage of a single star or a small number of stars has been proposed as a likely mechanism for producing large amplitude variability in AGNs \citep{lawrenceSlowblueNuclearHypervariables2016}. These events are uniform in color across the rest-frame UV and optical wavelengths, and the resulting light curves display symmetric profiles. The expected characteristic timescale is $O(\mathrm{yrs})$, determined by the lens mass and the relative transverse velocity between the source and the lens \citep{lawrenceSlowblueNuclearHypervariables2016}. Under the assumption that both the background source and the lens are point-like and their relative motion is linear, the time-dependent magnification can be approximated as \citep{grahamUnderstandingExtremeQuasar2017}:
\begin{equation}
  \label{eq:microlensing}
  A(t) = \frac{F_{\nu,obs} (t)}{F_{\nu,0}} = \frac{u(t)^{2} + 2}{u(t)\sqrt{u(t)^{2}+4}},
\end{equation}
where $u(t)^{2} = (t-t_{0})^{2}/t_{E}^{2} + u_{0}^{2}$ is the angular distance in units of the Einstein radius between the lens and the source \citep[for details see, e.g. ][]{schneiderGravitationalLenses1992} and $F_{\nu,obs}(t)$ and $F_{\nu,0}$ are the observed flux density and the baseline flux density.

\begin{figure*}
  \gridline{\fig{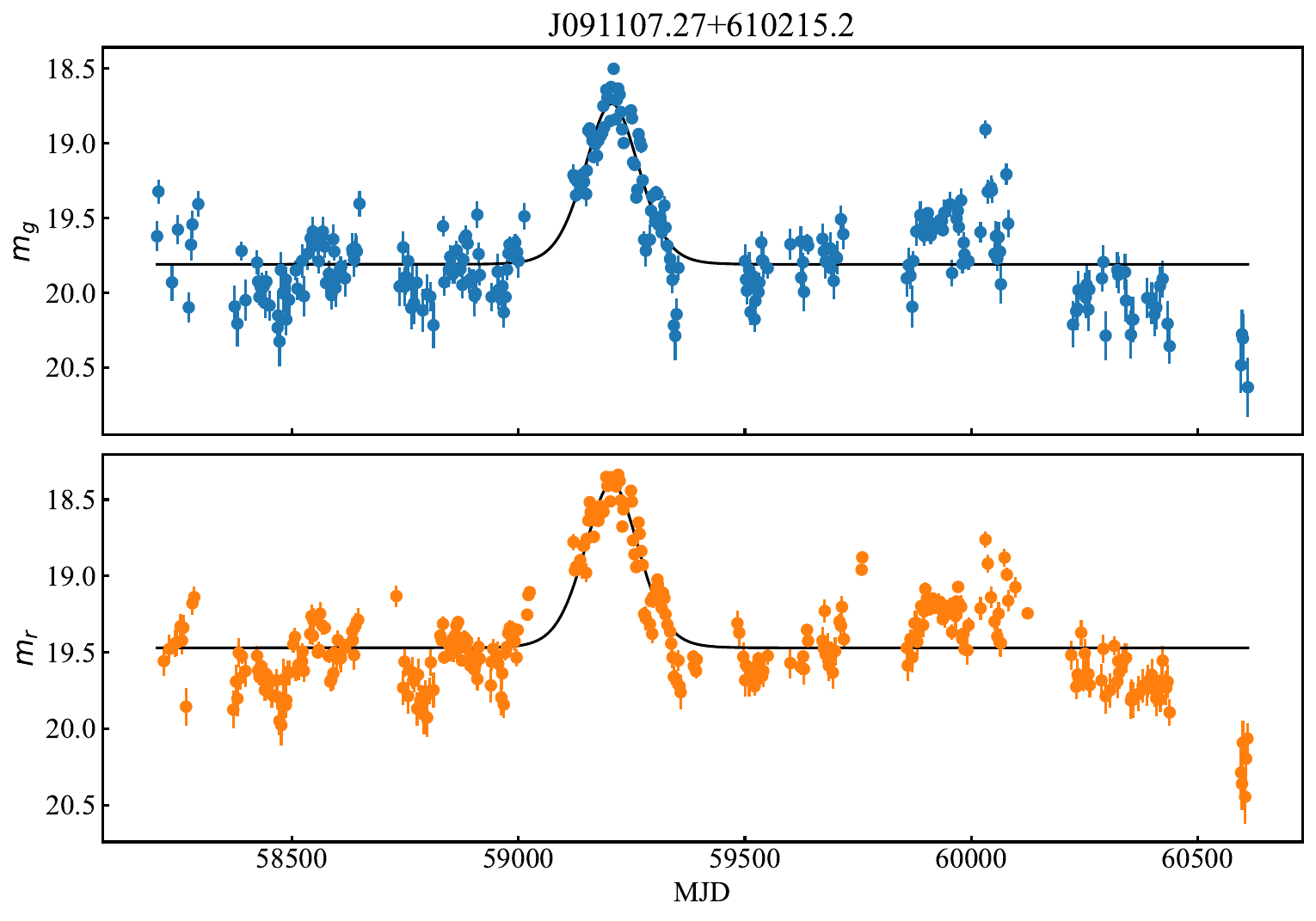}{0.5\textwidth}{}
  \fig{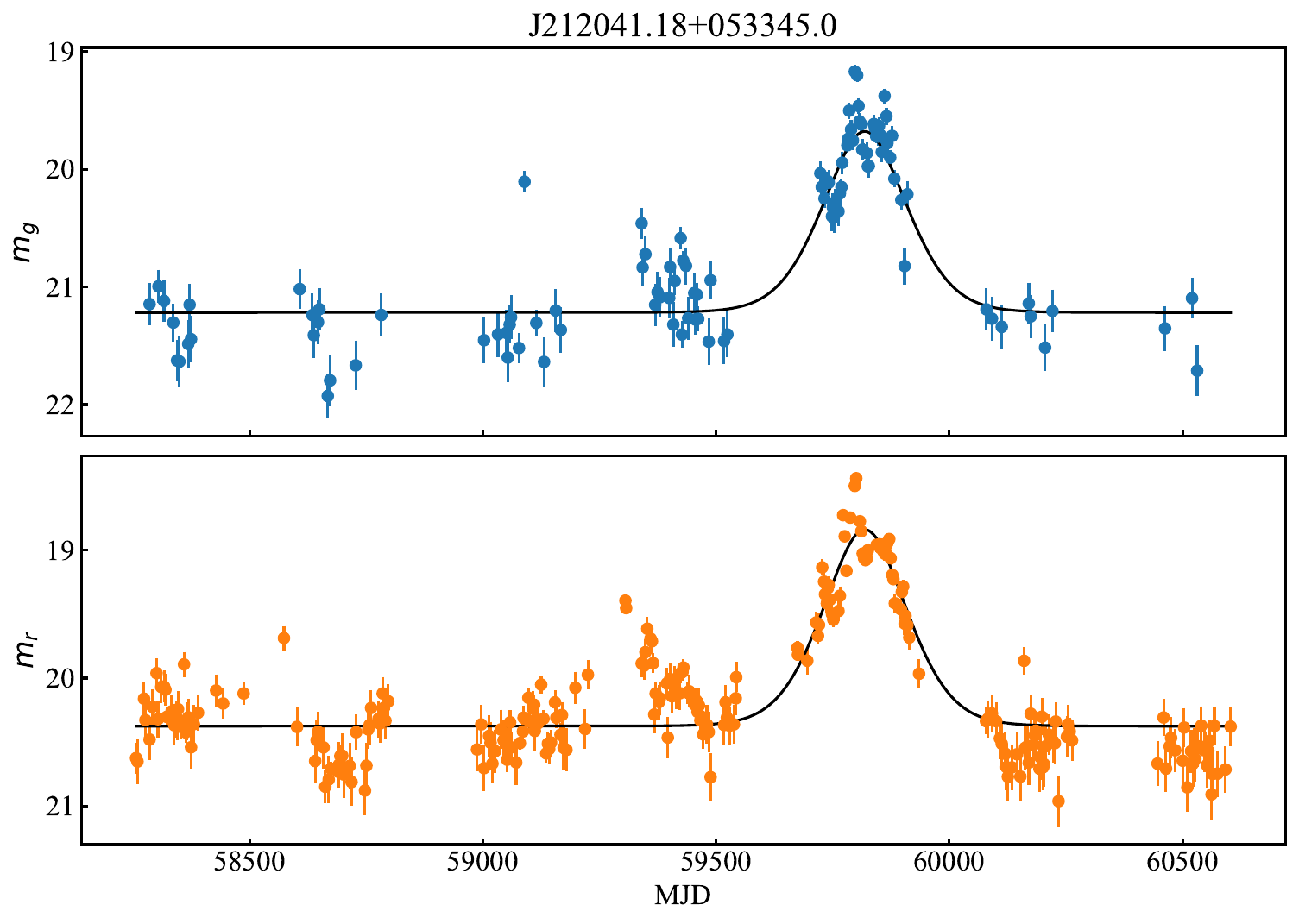}{0.5\textwidth}{}}
  \caption{Light curves of AGN flares that are better described by a symmetric microlensing profile than by the asymmetric flare profile. The solid lines show the best-fitting microlensing curves. }
  \label{fig:Lightcurve-microlensing}
\end{figure*}

We fit each light curve in the AGNFCC using the symmetric microlensing profile defined in Equation (\ref{eq:microlensing}), and compare the resulting BIC value to that obtained from the asymmetric profile defined in Equation (\ref{eq:flare}). A significantly lower BIC for the microlensing fit indicates a statistical preference for a lensing origin. Figure \ref{fig:Lightcurve-microlensing} presents two representative examples in which the light curves exhibit the characteristic symmetry expected from microlensing events.

\subsection{Blazars}

Blazars are a subclass of AGNs characterized by relativistic jets that are closely aligned with the line of sight to the observer. This alignment results in strong Doppler boosting, producing highly variable emission across the entire electromagnetic spectrum \citep{urryUnifiedSchemesRadioLoud1995}. Their optical light curves can exhibit large-amplitude, rapid, and erratic variability, with timescales ranging from minutes to years \citep{otero-santosOpticalVariabilityBlazar2024}.

\begin{figure*}
  \centering
  \gridline{\fig{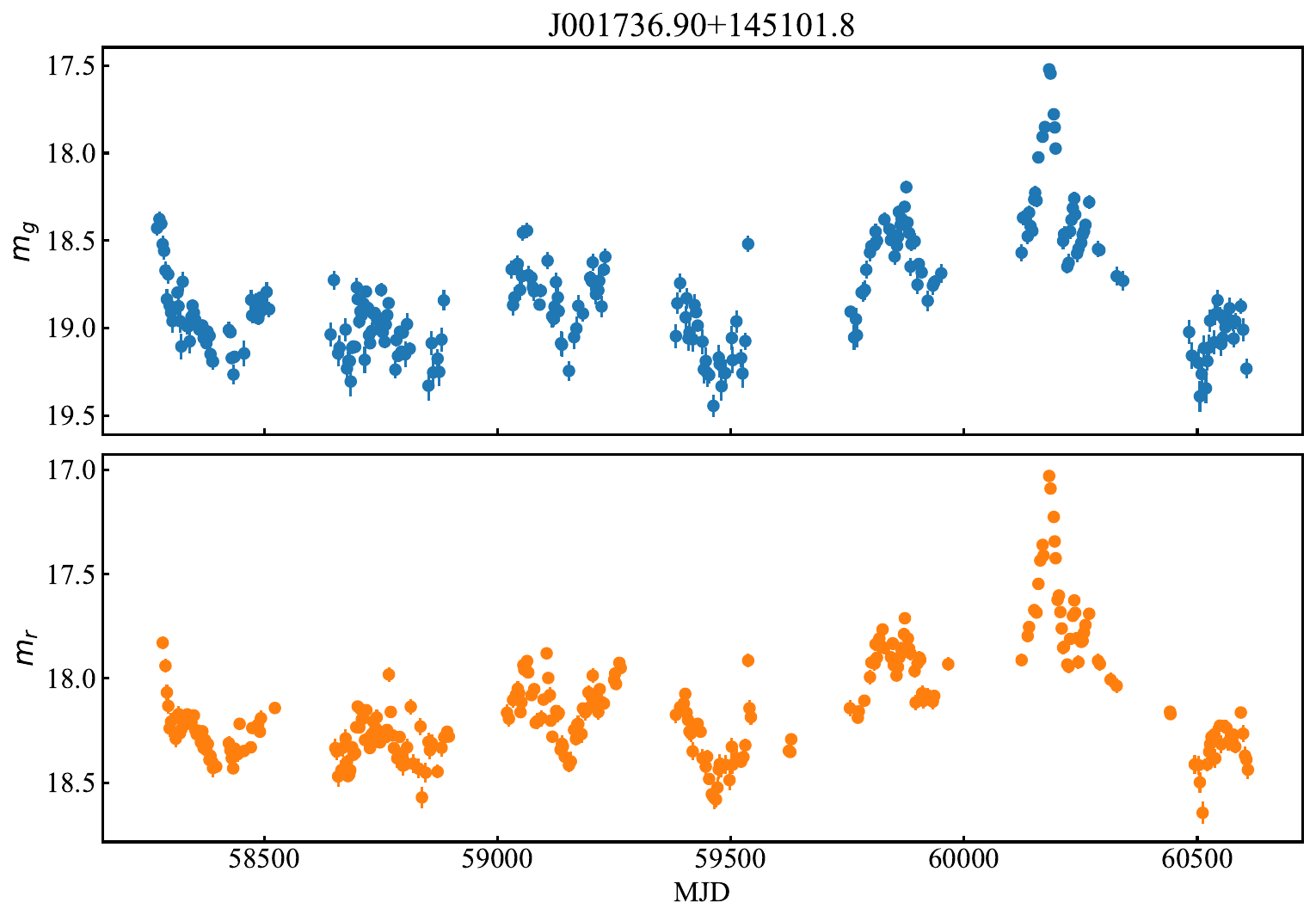}{0.5\textwidth}{}
    \fig{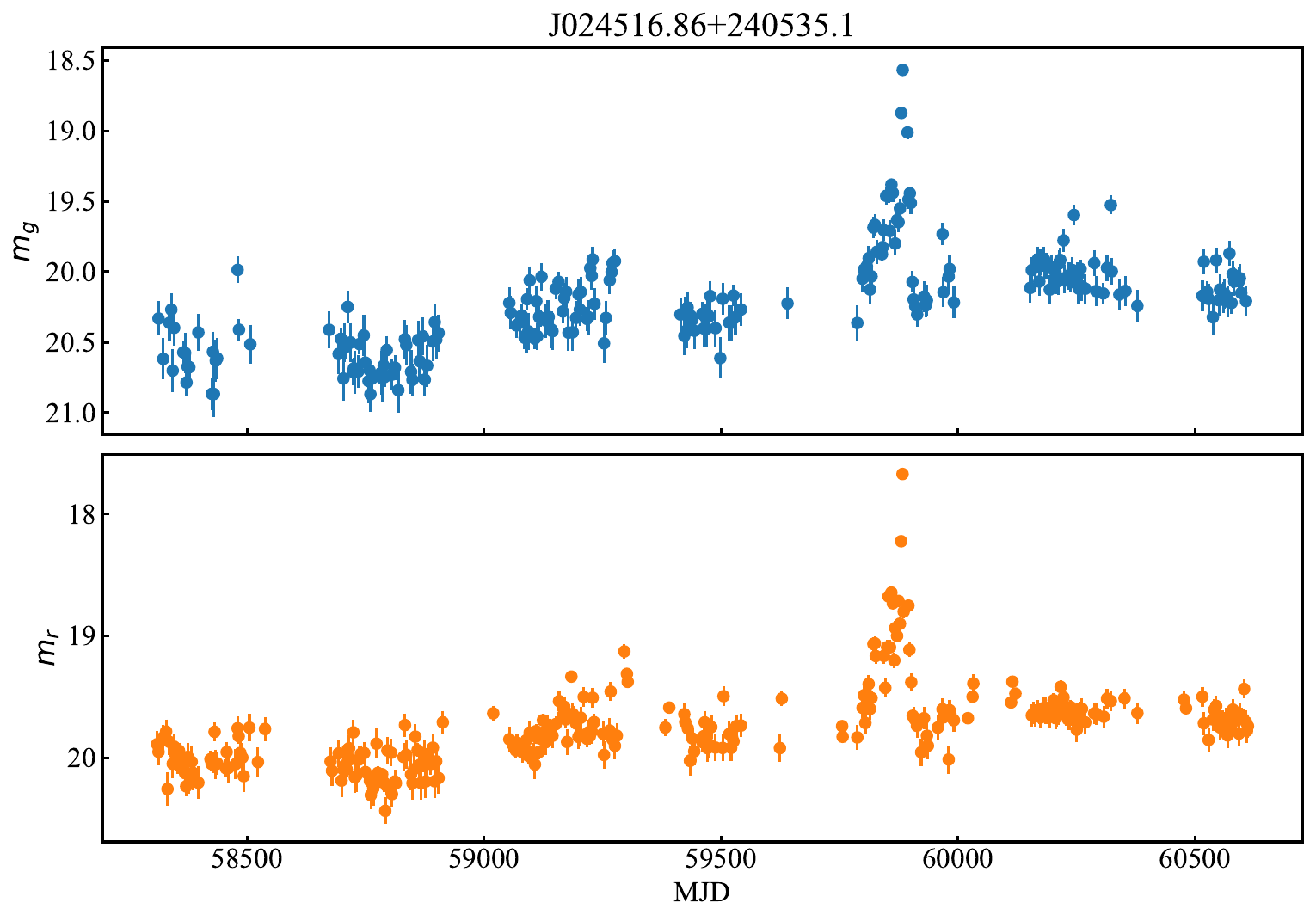}{0.5\textwidth}{}
  }
  \caption{Light curves of SDSS J001736.90+145101.8 (left) and B2 0242+23 (right), both are known blazars.}
  \label{fig:Lightcurve-blazar}
\end{figure*}

In our sample, blazars naturally account for a significant fraction of high-variability AGN flares. We find 830 blazars in the AGNFCC based on Roma-BZCAT and milliquas (see Section \ref{sec:SourceClassification}), representing approximately 3\% of the total flares. This fraction increases to 10\% in the AGNFRC, reflecting the strong variability that makes blazars more likely to pass our selection criteria. This trend is consistent with the changes shown in Figure \ref{fig:AGNType} and Figure \ref{fig:ZTFAlert}. Figure \ref{fig:Lightcurve-blazar} presents two representative examples.

\subsection{AGN Variability}
While our flare selection aims to identify events beyond the scope of AGN variability, it is important to recognize that intrinsic variability in AGNs cannot be entirely ruled out. According to our threshold of $p_{\mathrm{flare}} = 0.998$, at least 0.2\% of AGNs show variability strong enough to appear in our sample. Given the long observational baseline of ZTF, such rare but strong variability have a higher chance of being recorded, suggesting that this fraction is likely an underestimate. As illustrated in Figure \ref{fig:FlareProb}, a substantial number of AGNs exhibit $p_{\mathrm{flare}}$ values approaching this threshold.

\added{The non-flaring photometric variability of AGNs is widely characterized by a damped random walk \citep[DRW,][]{macleodMODELINGTIMEVARIABILITY2010} process. To evaluate the potential contamination introduced by this stochastic variability, we generated a set of 15,000 simulated AGN light curves with the DRW model, following the procedure described in \citep{mclaughlinUsingGaussianProcesses2024}. Each simulated light curve spans 1,000 days with a 3-day cadence and a photometric uncertainty of 0.1 mag, representing the typical variability behavior of AGNs under the assumption that genuine flares are intrinsically rare.

Applying our flare detection pipeline to these simulated light curves, we find that 61 sources($\sim 0.4\%$) pass the initial GP-based detection step. This rate is slightly higher than the nominal 0.2\% threshold but remains broadly consistent with our expectations. After applying the stricter selection criteria summarized in Table~\ref{tab:SelectionCriteria}, no source remains, highlighting the effectiveness of our pipeline design in minimizing false positives originating from the underlying DRW variability. It is worth noting that these simulations assume idealized DRW behavior and noiseless sampling. Real AGN light curves can exhibit deviations from the DRW process, coupled with instrumental and environmental effects, which may affect the actual false-positive rate in the observed data.}

The physical origin of such variability remains uncertain and has been debated for decades. Several theoretical mechanisms have been proposed, including the local \citep{lyubarskiiFlickerNoiseAccretion1997} or the global accretion rate \citep{liAccretionDiscModel2008} fluctuations, thermal fluctuations in the disk \citep{kellyACTIVEGALACTICNUCLEUS2013}, X-ray reprocessing \citep{kubotaPhysicalModelBroadband2018}, or the combination of these effects \citep{sanchez-saezQUESTSillaAGN2018}. In some cases, especially when the variability is strong or sparsely sampled, it may resemble a flare and satisfy our selection criteria.

Disentangling such intrinsic variability from others remains a challenge. \added{Alternative time-series frameworks, such as autoregressive moving average (ARMA) or Heterogeneous Autoregressive (HAR) models \citep{10.1093/jjfinec/nbp001,box2015time,chatfieldAnalysisTimeSeries2019,feigelsonAutoregressiveTimesSeries2018}, may provide useful diagnostics for assessing whether AGN variability can be fully described by stochastic processes or requires transient components, which we plan to investigate in future work.} Ultimately, moving beyond solely photometric modeling, spectroscopic and multi-wavelength observations will be essential for further refining flare selection and advancing our physical understanding of these events.

\section{Summary\label{sec:summary}}
In this work, we conduct a systematic search for AGN flares based on data from ZTF DR 23, one of the most extensive time-domain datasets available for the northern sky. Starting from a parent sample of approximately two million AGNs with sufficient g and r band observations, we construct two AGN flare catalogs: AGN Flare Coarse Catalog (AGNFCC), containing 28,504 AGN flares identified via Bayesian block segmentation and Gaussian Process modeling, and AGN Flare Refined Catalog (AGNFRC), containing 1,984 high-confidence flares selected by applying a set of strict quality and variability criteria.

We characterize the statistical properties of these AGN flares, including their spatial and redshift distributions, timescales, and host AGN types. The flare timescales span a wide range, from just a few days to several days, reflecting the diversity of physical mechanisms that may drive these events. The increase fraction of low redshift sources and the decreasing fraction of QSOs from the general AGN sample to the flare catalogs suggests that flares are more readily detectable in lower-luminosity systems.

We further investigate the possible origins of the detected flares. By cross-matching with external catalogs, we identify subsets of flares associated with known supernovae, TDEs, blazars, and candidate electromagnetic counterparts to BBH mergers. In addition, several cases show evidence consistent with microlensing. Nevertheless, a large fraction of flares remains unexplained and may reflect the intrinsic variability of AGNs.

The AGNFCC and AGNFRC provide a valuable foundation for future studies of the transient phenomena in AGNs. While the AGNFRC represents a high-confidence sample, some events may still not correspond to true flares due to complex AGN variability or sparse sampling. Additional spectroscopic and multi-wavelength observations will therefore be crucial for refining flare selection and classification, understanding their physical mechanism, and exploring their potential connections to AGN disks.

%%%%%%%%%%%%%%%%%%%%%%%%%%%%%%%%%%%%%%%%%%%%%%%%%%%%%%%%%%%

\begin{acknowledgements}

  This work is supported by Strategic Priority Research Program of the Chinese Academy of Science (Grant No. XDB0550300), the National Key R\&D Program of China (Grant No. 2021YFC2203102, 2022YFC2204602, 2024YFC2207500, and 2022YFC2807303), the National Natural Science Foundation of China (Grant No. 12325301, 12273035, and 12405075), the Science Research Grants from the China Manned Space Project (Grant No. CMS-CSST-2021-B01), the 111 Project for ``Observational and Theoretical Research on Dark Matter and Dark Energy" (Grant No. B23042), and Cyrus Chun Ying Tang Foundations.

This work is based on observations obtained with the Samuel Oschin Telescope 48-inch and the 60-inch Telescope at the Palomar Observatory as part of the Zwicky Transient Facility project. ZTF is supported by the National Science Foundation under Grants No. AST-1440341 and AST-2034437 and a collaboration including current partners Caltech, IPAC, the Oskar Klein Center at Stockholm University, the University of Maryland, University of California, Berkeley , the University of Wisconsin at Milwaukee, University of Warwick, Ruhr University, Cornell University, Northwestern University and Drexel University. Operations are conducted by COO, IPAC, and UW.
\end{acknowledgements}

\software{
  celerite \citep{foreman-mackeyFastScalableGaussian2017}, astropy \citep{astropycollaborationAstropyCommunityPython2013, astropycollaborationAstropyProjectBuilding2018, astropycollaborationAstropyProjectSustaining2022}, matplotlib \citep{hunterMatplotlib2DGraphics2007}, numpy \citep{harrisArrayProgrammingNumPy2020}, seaborn \citep{Waskom2021}, pandas \citep{mckinney-proc-scipy-2010}
 }

\appendix
\section{Light Curve Examples of Removed Flares for AGNFRC \label{appendix:excluded_lcs}}
\begin{figure}[htb!]

  \gridline{
    \includegraphics[trim=10 10 0 8, clip,width=0.45\textwidth]{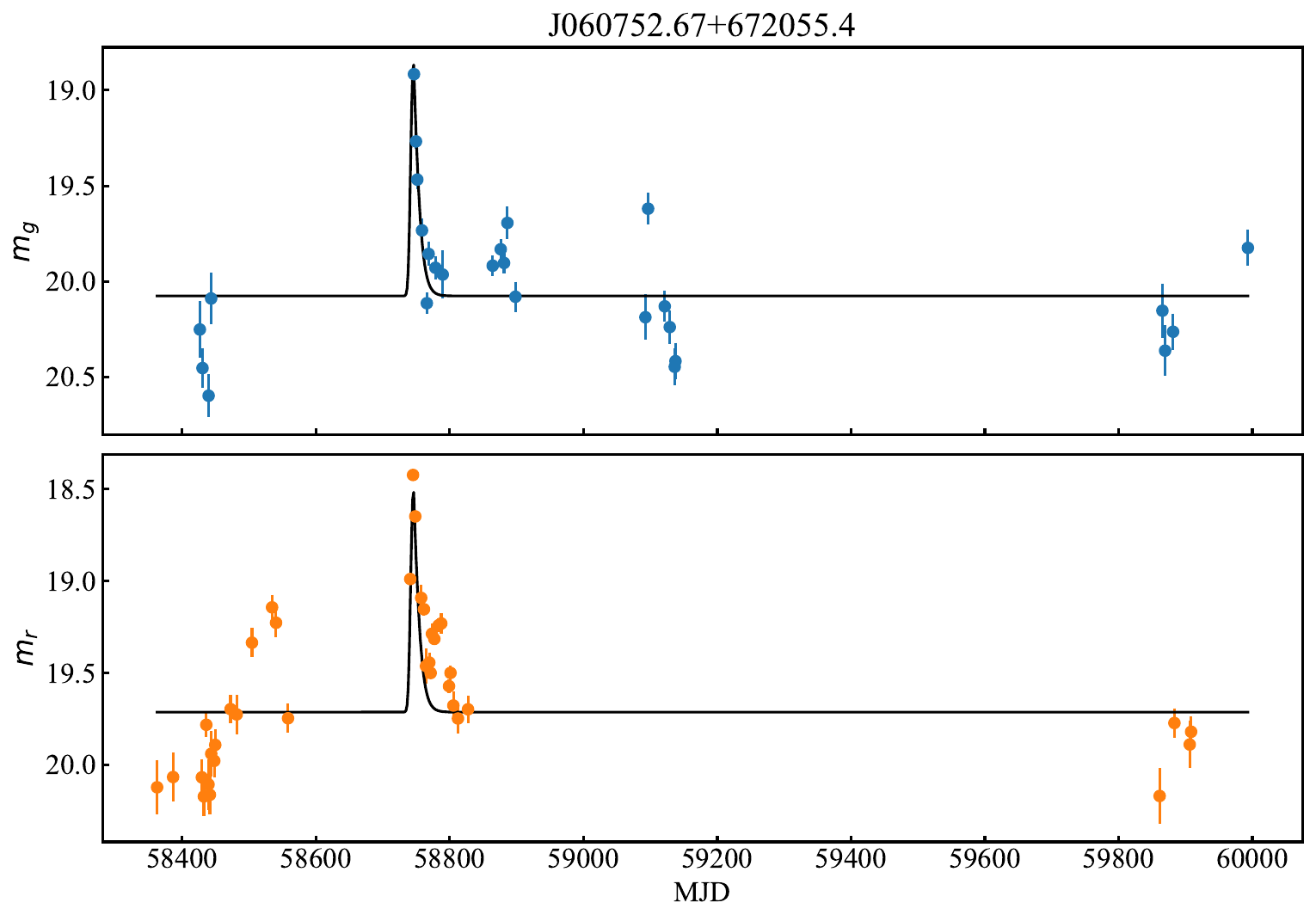}
    \includegraphics[trim=10 10 0 8, clip,width=0.45\textwidth]{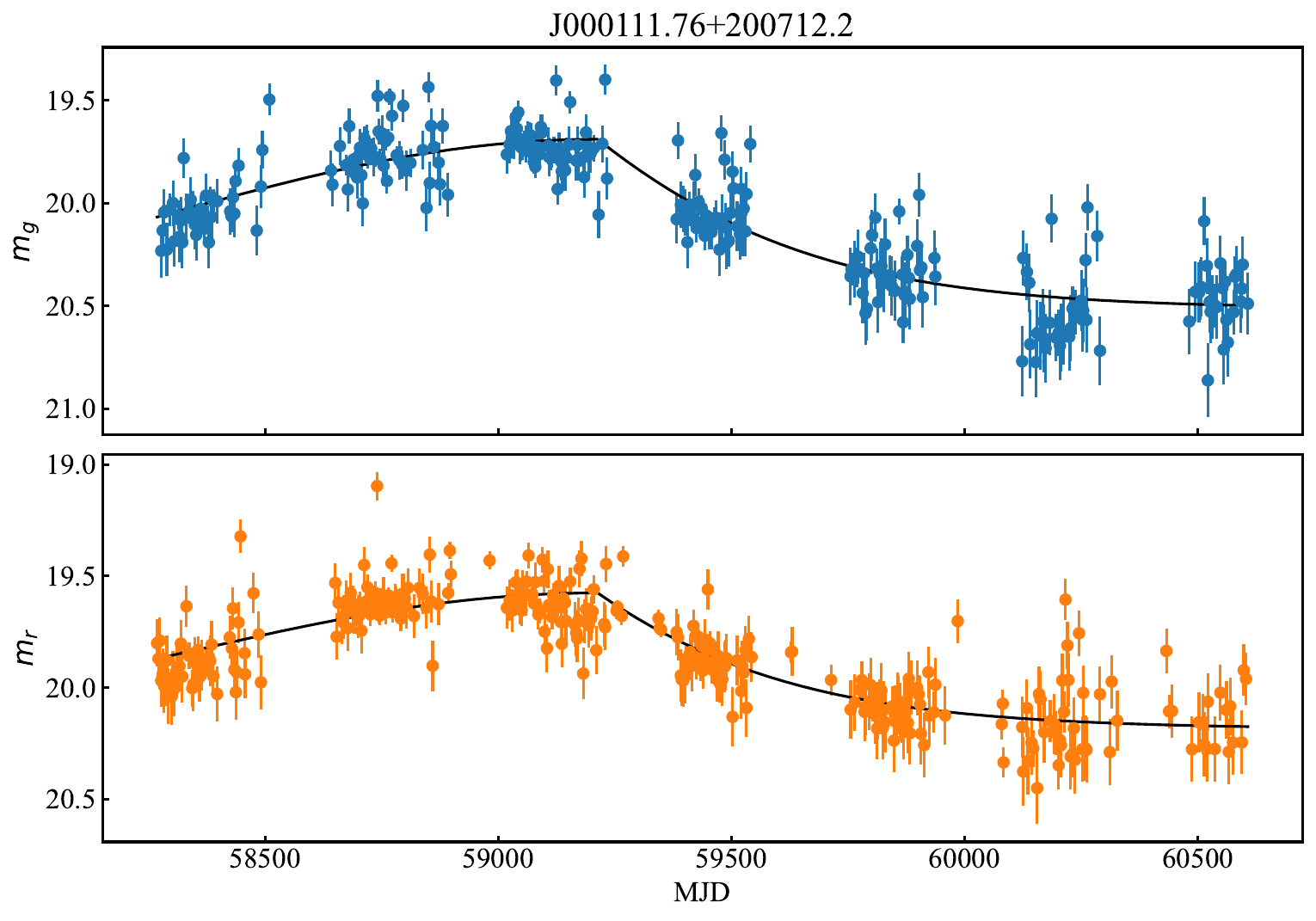}
  }
  \gridline{
    \includegraphics[trim=10 10 0 8, clip,width=0.45\textwidth]{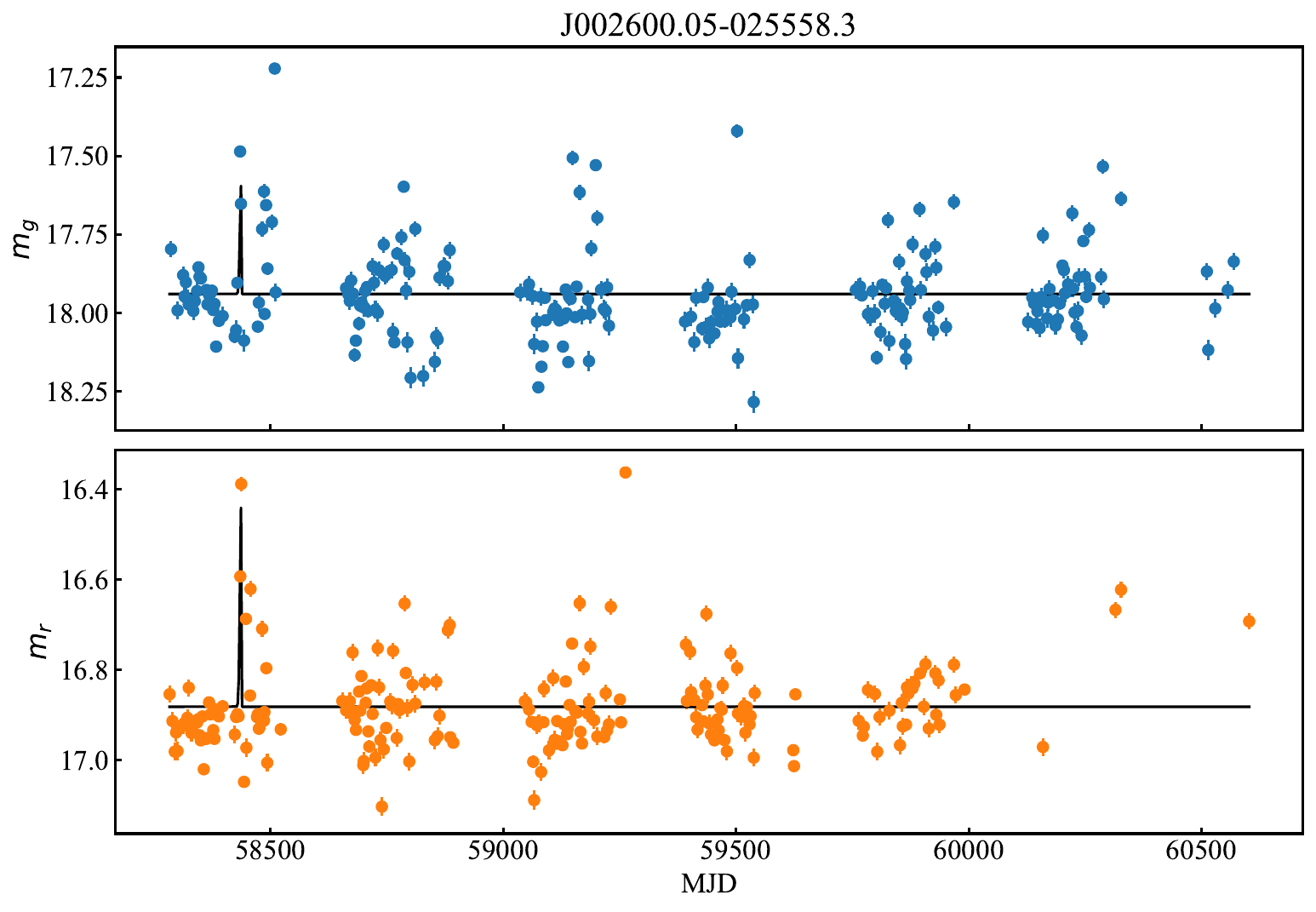}
    \includegraphics[trim=10 10 0 8, clip,width=0.45\textwidth]{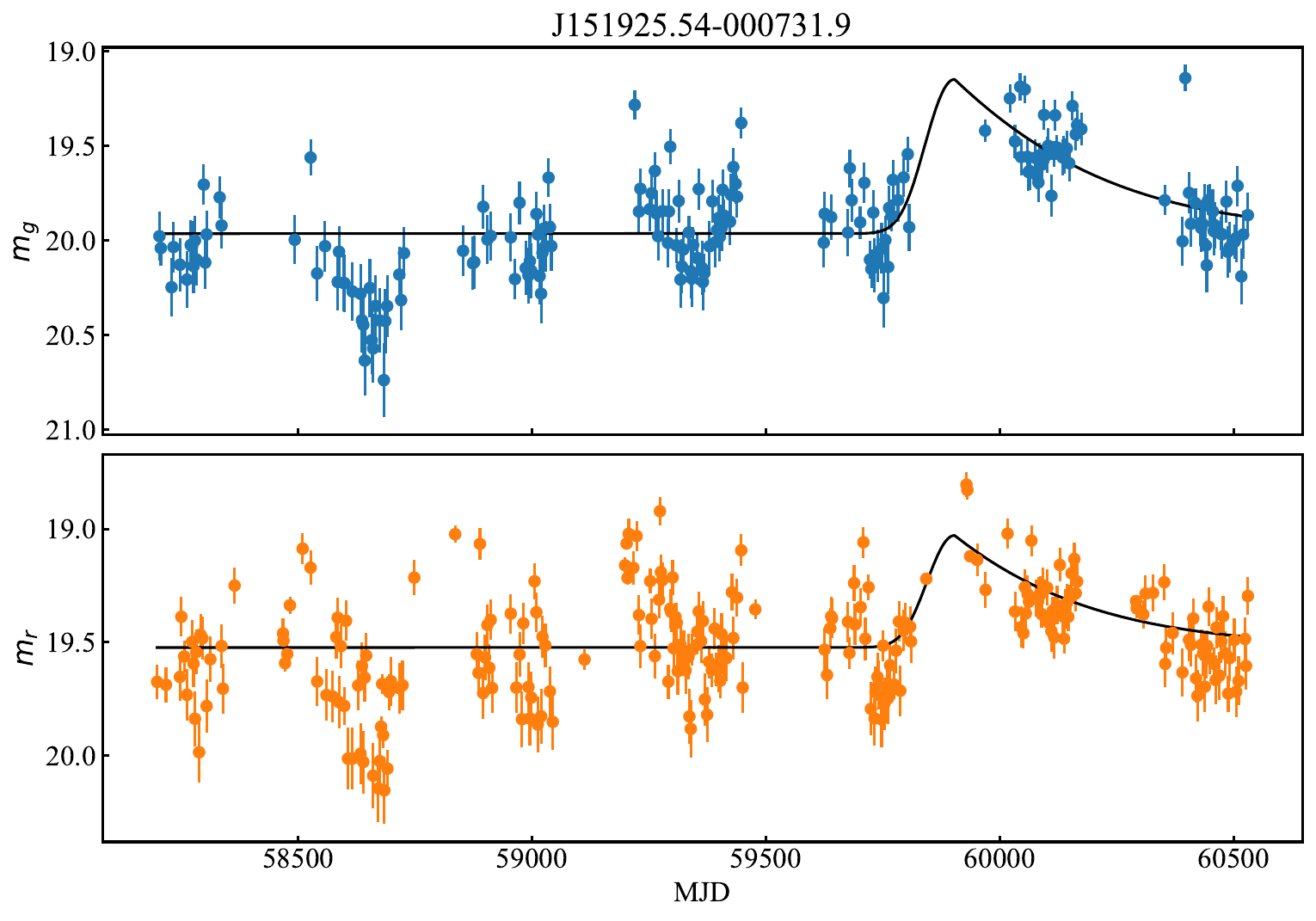}
  }
  \gridline{
    \includegraphics[trim=10 10 0 8, clip,width=0.45\textwidth]{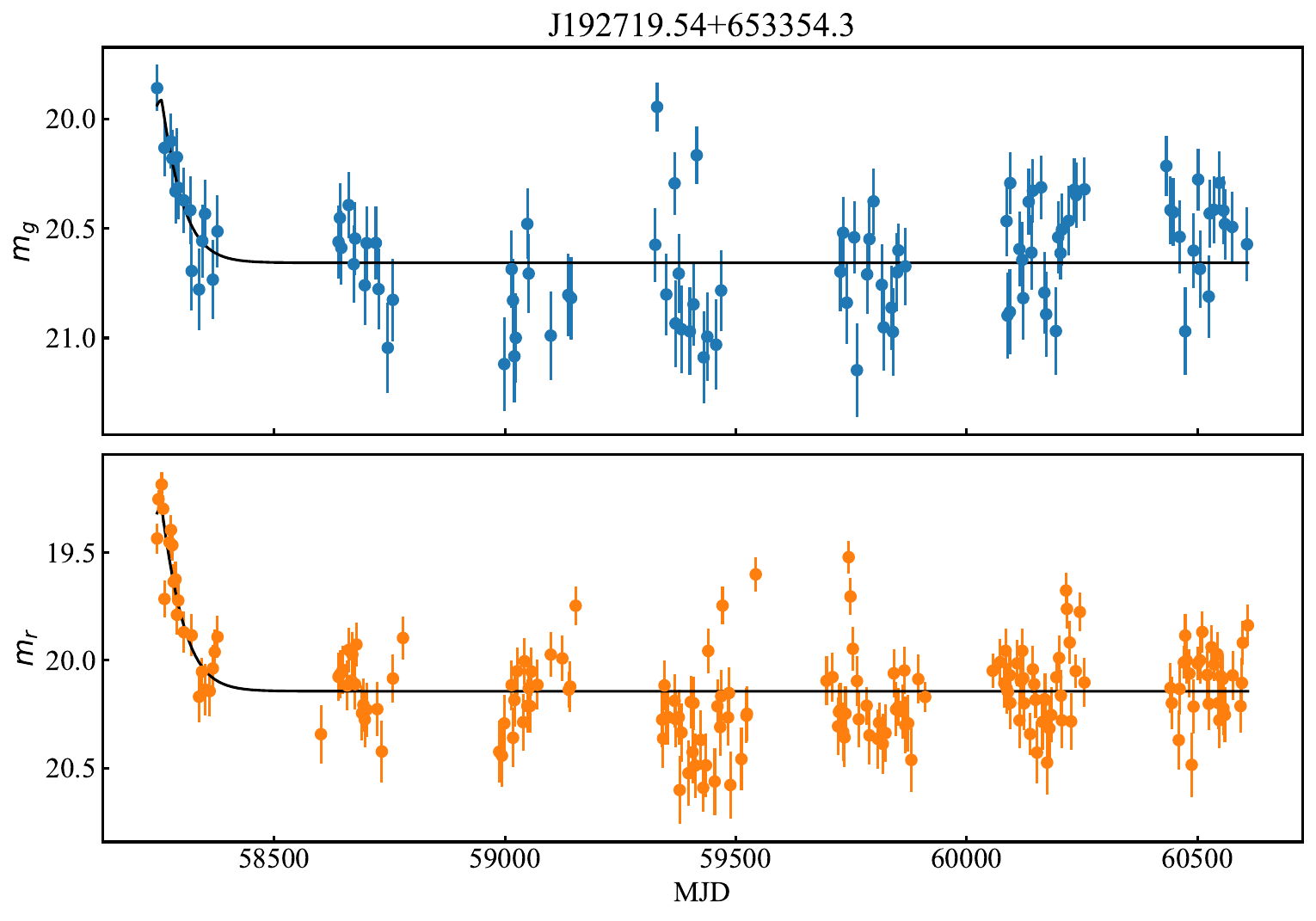}
    \includegraphics[trim=10 10 0 8, clip,width=0.45\textwidth]{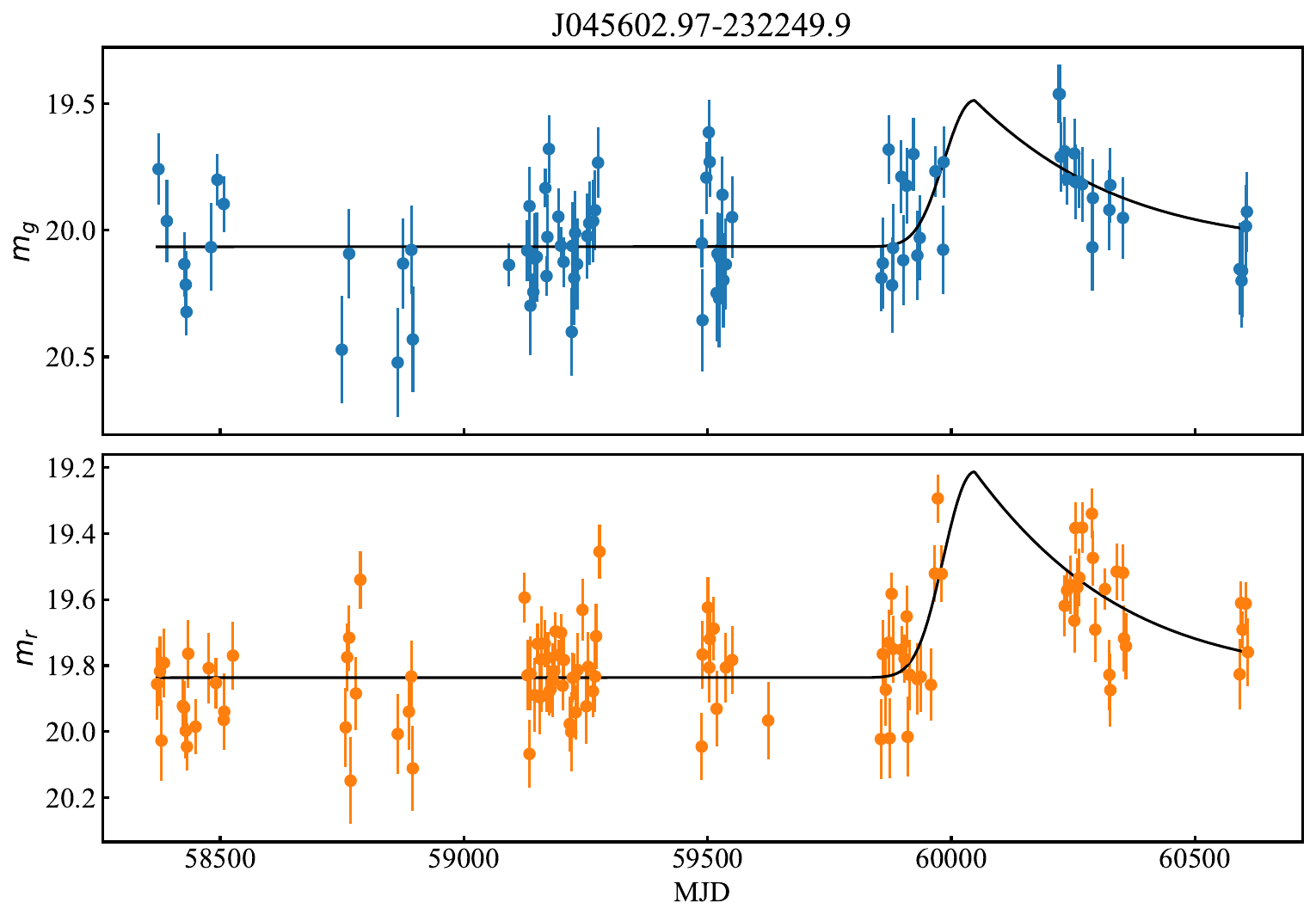}
  }
  \gridline{
    \includegraphics[trim=10 10 0 8, clip,width=0.45\textwidth]{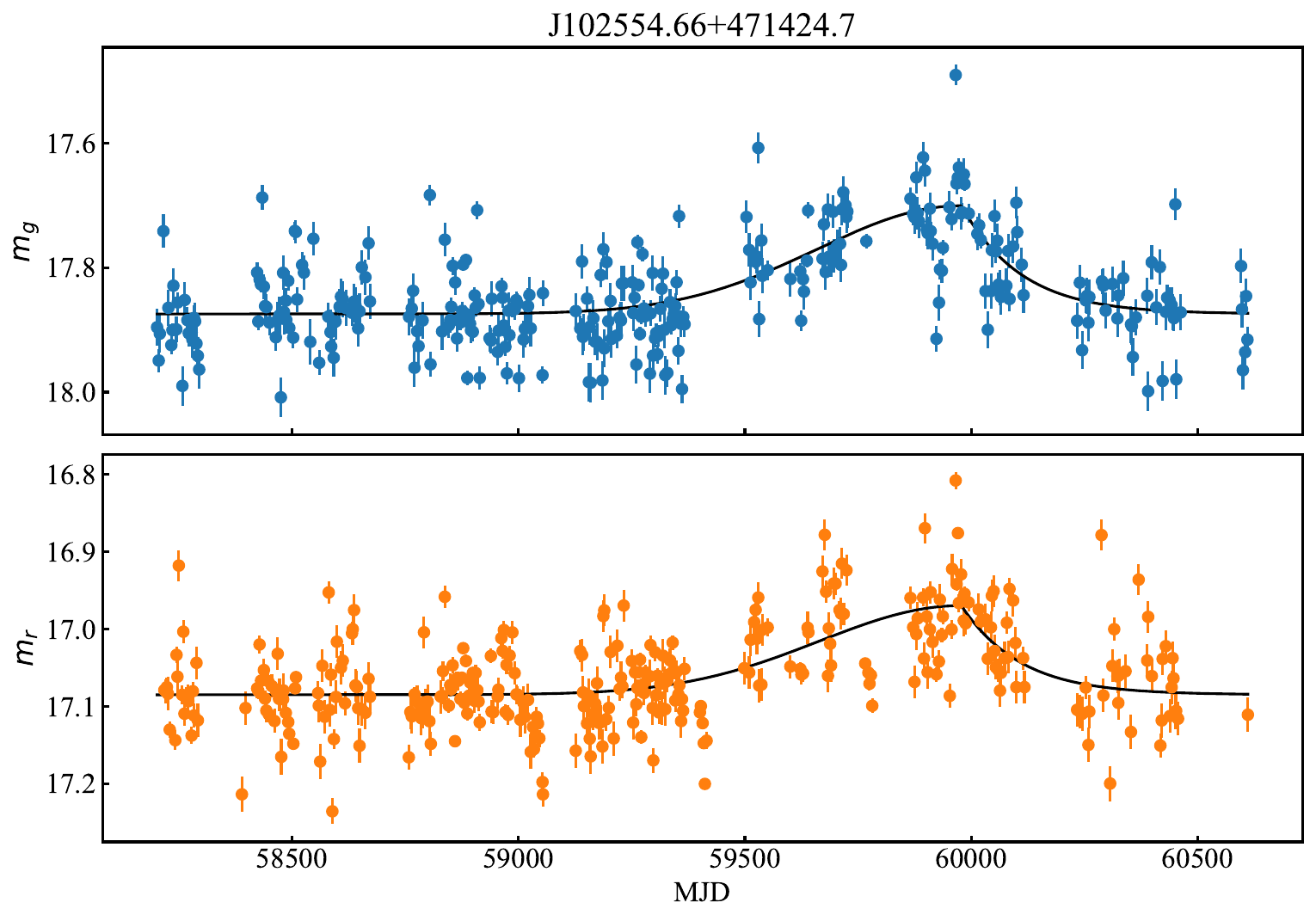}
    \includegraphics[trim=10 10 0 8, clip,width=0.45\textwidth]{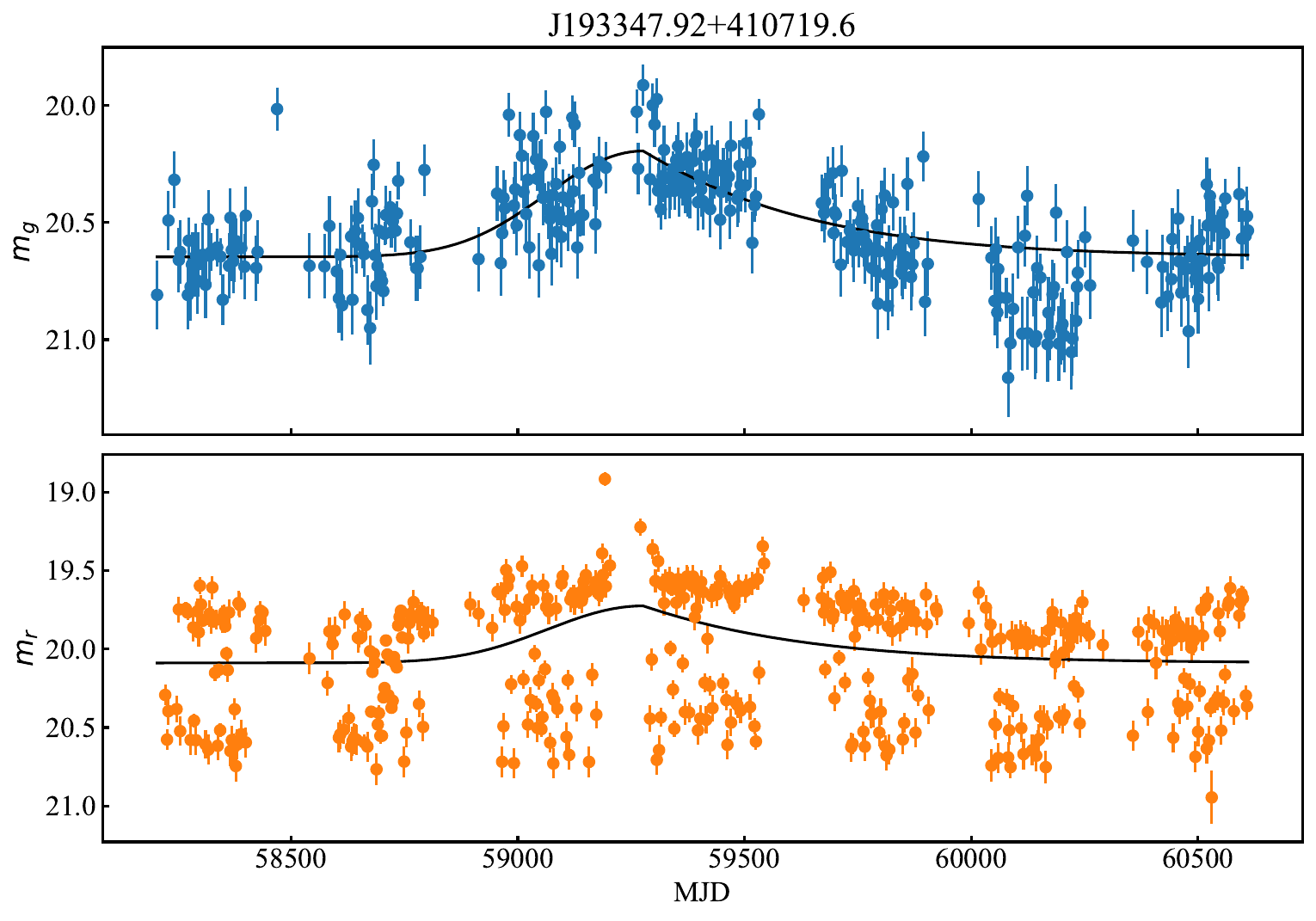}
   }

  \caption{Light curves of eight flares in the AGNFCC that are excluded from the AGNFRC, each corresponding to one of the eight selection criteria listed in Table \ref{tab:SelectionCriteria}. Panels are arranged from left to right, top to bottom, in the order of criterion 1 through 8. The black lines indicate the best-fitting profiles using Equation (\ref{eq:flare}).}
  \label{fig:Lightcurve-excluded}
\end{figure}

Here, we present representative light curves of flares that are excluded from the AGNFRC due to not meeting one or more of the refined selection criteria in Table \ref{tab:SelectionCriteria}. These examples illustrate common failture modes, such as insufficient temporal coverage, low variability significance, or contamination from nearby sources. They help to highlight the importance and the effectiveness of the filtering conditions applied in constructing a high-significance flare catalog. These light curves are shown in Figure \ref{fig:Lightcurve-excluded}.

\bibliography{main}
\bibliographystyle{aasjournalv7}
\end{document}